\documentclass[twocolumn,dvipsnames,twocolappendix]{aastex631}
\usepackage{amsmath,amsfonts,amssymb,graphicx,chngcntr,multirow, float,booktabs}
\usepackage[]{hyperref}
\hypersetup{colorlinks=true}

\received{Dec. 19, 2024}
\revised{Feb. 1, 2025}
\accepted{Feb. 21, 2025}

\shorttitle{QPE spectra and homologous expansion}
\shortauthors{Chakraborty \textit{et al.}}

\begin{document}

\title{Rapidly varying ionization features in a Quasi-periodic Eruption: a homologous expansion model for the spectroscopic evolution}

\author[0000-0002-0568-6000]{Joheen Chakraborty}\thanks{joheen@mit.edu}
\affiliation{Department of Physics \& Kavli Institute for Astrophysics and Space Research, Massachusetts Institute of Technology, Cambridge, MA 02139, USA}

\author[0000-0003-4511-8427]{Peter Kosec}\thanks{NASA Einstein Fellow}
\affiliation{Center for Astrophysics | Harvard \& Smithsonian, Cambridge, MA, USA}

\author[0000-0003-0172-0854]{Erin Kara}
\affiliation{Department of Physics \& Kavli Institute for Astrophysics and Space Research, Massachusetts Institute of Technology, Cambridge, MA 02139, USA}

\author[0000-0003-0707-4531]{Giovanni Miniutti}
\affiliation{Centro de Astrobiolog\'ia (CAB), CSIC-INTA, Camino Bajo del Castillo s/n, 28692 Villanueva de la Ca\~nada, Madrid, Spain}

\author[0000-0003-4054-7978]{Riccardo Arcodia}\thanks{NASA Einstein Fellow}
\affiliation{Department of Physics \& Kavli Institute for Astrophysics and Space Research, Massachusetts Institute of Technology, Cambridge, MA 02139, USA}

\author[0000-0001-9735-4873]{Ehud Behar}
\affiliation{Department of Physics \& Kavli Institute for Astrophysics and Space Research, Massachusetts Institute of Technology, Cambridge, MA 02139, USA}
\affiliation{Department of Physics, Technion, Haifa 32000, Israel}

\author[0000-0002-1329-658X]{Margherita Giustini}
\affiliation{Centro de Astrobiolog\'ia (CAB), CSIC-INTA, Camino Bajo del Castillo s/n, 28692 Villanueva de la Ca\~nada, Madrid, Spain}

\author[0000-0002-8606-6961]{Lorena Hernández-García}
\affiliation{Millennium Nucleus on Transversal Research and Technology to Explore Supermassive Black Holes (TITANS)}
\affiliation{Millennium Institute of Astrophysics (MAS), Nuncio Monseñor Sótero Sanz 100, Providencia, Santiago, Chile}
\affiliation{Instituto de F\'isica y Astronom\'ia, Facultad de Ciencias, Universidad de Valpara\'iso, Gran Breta\~na 1111, Playa Ancha, Valpara\'iso, Chile}

\author[0000-0003-4127-0739]{Megan Masterson}
\affiliation{Department of Physics \& Kavli Institute for Astrophysics and Space Research, Massachusetts Institute of Technology, Cambridge, MA 02139, USA}

\author[0000-0002-7116-2897]{Erwan Quintin}
\affiliation{European Space Agency (ESA), European Space Astronomy Centre (ESAC), Camino Bajo del Castillo s/n, 28692 Villanueva de la Cañada, Madrid, Spain}

\author[0000-0001-5231-2645]{Claudio Ricci}
\affiliation{Instituto de Estudios Astrofísicos, Facultad de Ingeniería y Ciencias, Universidad Diego Portales, Avenida Ejercito Libertador 441, Santiago, Chile}
\affiliation{Kavli Institute for Astronomy and Astrophysics, Peking University, Beijing 100871, China}

\author[0000-0003-0820-4692]{Paula Sánchez-Sáez}
\affiliation{European Southern Observatory, Karl-Schwarzschild-Strasse 2, 85748 Garching bei München, Germany}

\begin{abstract}
Quasi-Periodic Eruptions (QPEs) are recurring bursts of soft X-ray emission from supermassive black holes (SMBHs), which a growing class of models explains via extreme mass-ratio inspirals (EMRIs). QPEs exhibit blackbody-like emission with significant temperature evolution, but the minimal information content of their almost pure-thermal spectra has limited physical constraints. Here we study the recently discovered QPEs in ZTF19acnskyy (``Ansky''), which show absorption-like features evolving dramatically within eruptions and correlating strongly with continuum temperature and luminosity, further probing the conditions underlying the emission surface. The absorption features are well-described by dense ionized plasma of column density $N_{\rm H}\gtrsim 10^{21}$ cm$^{-2}$, blueshift $0.06\lesssim v/c \lesssim 0.4$, and either collisional or photoionization equilibrium. With high-resolution spectra, we also detect ionized blueshifted emission lines suggesting a nitrogen over-abundance of $21.7^{+18.5}_{-11.0}\times$ solar. We interpret our results with orbiter-disk collisions in an EMRI system, in which each impact drives a shock that locally heats the disk and expels X-ray emitting debris undergoing radiation pressure-driven homologous expansion. We explore an analytical toy model that links the rapid change in absorption lines to the evolution of the ionization parameter and the photosphere radius, and suggest that $\sim 10^{-3}M_\odot$ ejected per eruption with expansion velocities up to $v_{\rm max}\sim 0.15c$, can reproduce the absorption features. With these assumptions, we show a P Cygni profile in a spherical expansion geometry qualitatively matches the observed line profiles. Our work takes a first step towards extending existing physical models for QPEs to address their implications for spectral line formation.
\end{abstract}

\keywords{}

\section{Introduction} \label{sec:intro}
Quasi-Periodic Eruptions (QPEs) are a newly discovered, rapidly growing class of recurring X-ray transient from supermassive black holes (SMBHs) in nearby low-mass galaxy nuclei. Thus far, they have been observed with peak luminosities of  $L\sim 10^{42-44}$ erg s$^{-1}$, recurrence times of $P\sim 2.5-100\;$hr, blackbody-like spectra with temperatures of $kT\sim 50-250\,$ eV, SMBH masses of $\sim 10^{5-7.5}\;M_\odot$, and out to redshifts of $z\sim 0.08$. Since their discovery in 2019, the QPE population has grown to ten \citep{Miniutti19,Giustini20,Arcodia21,Arcodia24a,Chakraborty21,Chakraborty25,Quintin23,Nicholl24,Hernandez25}. A handful of potentially related phenomena are known, such as quasi-periodic oscillations in SMBHs \citep{Gierlinski08,Terashima12,Lin13,Pasham19,Masterson25} and recurring nuclear transients seen in optical \citep{Payne21,Somalwar23,Veres24} and X-ray \citep{Liu23,Wevers23,Evans23,Guolo24} wavelengths.

The serendipitous discovery of QPEs has inspired significant efforts to develop explanatory physical models, which separate broadly into two classes: (a) cyclic behavior related to thermal-viscous instabilities or tearing/precession of the SMBH accretion disk \citep{Raj21,Pan22,Pan23,Kaur23,Sniegowska23,Middleton25}; or (b) interaction of the SMBH (or its accretion disk) with a $\sim$stellar-mass orbiting companion in an extreme mass-ratio inspiral (EMRI; \citealt{Dai10,Xian21,Sukova21,King22,Zhao22,Lu23,Franchini23,Linial23b,Tagawa23,Kejriwal24,Zhou24a}). The latter class of models has garnered particular attention: if QPEs are indeed the electromagnetic counterparts of EMRIs, they are the first observational probe of EMRI rates \citep{Arcodia24b} and stellar dynamical phase space in external galaxy nuclei \citep{Rom24}, and QPEs themselves may eventually be multi-messenger sources and/or form a millihertz gravitational wave background \citep{Chen22} for upcoming space-based detectors (e.g. \textit{LISA}, \citealt{Amaro-Seoane17}; TianQin, \citealt{Luo16}).

Within the EMRI paradigm, recent activity has shifted away from scenarios which produce X-ray emission by direct accretion from the orbiting companion (e.g. \citealt{Metzger22,King22,Linial23a}), instead converging towards models which produce the emission via collisional shocks of the star or black hole companion with the underlying SMBH accretion disk (e.g. \citealt{Dai10,Xian21,Sukova21,Franchini23,Linial23b,Tagawa23,Zhou24a,Yao25}), which we refer to hereafter as ``orbiter-disk collision models''. There are several reasons to favor these models, of which we list three. 

\textbf{(1)} The viscous timescale at pericenter is uncomfortably long compared the several-hour timescales over which QPEs radiate their energy. Only highly eccentric EMRIs can achieve $t_{\rm visc}$ short enough to power QPEs via accretion, but they face a fine-tuning problem to produce orbits precisely constructed to overflow \textit{some} mass while avoiding tidal disruption. Furthermore, mass-overflow from a highly eccentric EMRI should produce strictly periodic events on a timescale short enough that secular effects (e.g. orbital hardening) can be neglected, which is at odds with the range of timing behaviors exhibited in QPEs. At the same time, QPE-emitting galaxies host compact accretion disks consistent with $L_X \propto T^4$, indicating there is a stable accretion flow whose accretion rate is linked to the eruption strength without powering them altogether \citep{Miniutti23a,Miniutti23b}. Producing the emission via collisional shocks avoids the timescale problem, while still linking QPEs to the disk properties.

\textbf{(2)} EMRI models elegantly explain the diverse range of irregularity and evolution in the burst times-of-arrival (i.e. the ``Q" in QPE). Some sources show a regular alternation between two recurrence times differing by 5-10\% \citep{Arcodia22,Miniutti23a}, which is naturally produced by a mildly eccentric EMRI colliding twice per orbit through ascending/descending node. Other sources show a more chaotic spread in recurrence times \citep{Arcodia22,Giustini24}, which have been found to show sinusoidal modulation on several-day super-periods \citep{Chakraborty24,Miniutti25} compatible with models of rigidly precessing accretion disks \citep{Franchini16,Franchini23}. In two cases, QPEs tentatively show evidence for a \textit{secular period decrease}, inferred from multi-year timing data of eRO-QPE2 \cite{Arcodia24c,Pasham24c} and GSN 069 \cite{Miniutti25}. This secular timing drift is a key prediction linked to the EMRI orbital decay via gravitational wave emission and viscous drag from the accretion disk \citep{Linial24a,linial24c}.

\textbf{(3)} QPE spectral evolution is atypical compared to other nuclear transients (e.g. tidal disruption events): QPEs show hysteresis in the $L-kT$ plane, i.e. harder/hotter bands peak before soft. For blackbody-like spectra, this can be used to infer a growing emission radius with time \citep{Miniutti23a,Chakraborty24,Arcodia24b}. While shock-driven EMRI models do not \textit{uniquely} explain this behavior, they provide a particularly natural physical explanation for it, best understood by analogy to the similar color-dependence in multi-band light curves of supernovae (e.g. Fig. 4 of \citealt{Kasen06}). Supernovae, like QPEs, are also thought to be powered by radiation emerging from a quasi-spherical region undergoing rapid homologous expansion following a shock breakout. Early after breakout, only the highest-energy photons can penetrate the dense, newly-compressed material. As photons continue to diffuse out, they lose energy to adiabatic losses as they do work expanding the ejecta. In this process, the ejecta become less dense and less optically thick, allowing the bulk of the photons to eventually escape by peak light. As a result, photons escaping earlier form a higher-$kT$ spectrum (hard bands peak early and quick), and over time $kT$ diminishes (soft bands peak late and slow). Radiative transfer simulations tracing the evolution of synthetic light curves and spectra following a orbiter-disk collision indeed find good agreement with observations of QPEs \citep{Vurm24}.

Direct tests of EMRI models are still in their infancy. Thus far, they have primarily focused on points \textbf{(1)} and \textbf{(2)} above, via modeling QPE timings with the orbital/precessional modes and evolution of the putative EMRIs \citep{Chakraborty24,Arcodia24c,Miniutti25,Pasham24b,Pasham24c,Zhou24a,Zhou24b} and fitting the properties of the underlying accretion disk to determine their compatibility with various model predictions \citep{Nicholl24,Guolo25,Wevers25}. Direct tests of the QPE radiation mechanism---point \textbf{(3)}---are less studied in comparison. They are complicated dramatically by the limited information content of the almost pure-thermal spectra of QPEs, making it difficult to probe the physical properties of the emission surface (e.g. expansion velocity, density, composition, ionization state). Inferring these would require detectable absorption/emission lines in addition to the continuum. This is observationally difficult, as most QPEs last only hours and have peak fluxes of $\sim 10^{-13}-10^{-12}$ erg cm$^{-2}$ s$^{-1}$; one simply cannot collect enough photons to confidently detect anything \textit{but} the continuum. The exception to this is the best-studied QPE source GSN 069, which shows evidence for persistent ionized absorption in both moderate-resolution \citep{Miniutti23a} and high-resolution spectra \citep{Kosec25}. The latter study, by stacking $>1$ Ms of \textit{XMM-Newton} RGS data, found the first evidence for blueshifted absorption at a velocity of $\sim 0.01c$. The long integrations needed to generate enough signal for these detections meant they could not be probed in a phase-resolved manner, instead averaging over many bursts.

The recent discovery of QPEs in ZTF19acnskyy (``Ansky'';  \citealt{Hernandez25}), a $10^6\;M_\odot$ SMBH at redshift $z=0.024$ with a newborn accretion flow \citep{Sanchez24}, provides an exciting opportunity to overcome the observational barriers laid out above. With the longest burst durations/peak fluxes by factors of $4/2.5\times$ respectively, Ansky accumulates 10$\times$ more counts per eruption than any other known QPE source. The high flux is due to the relatively large peak luminosity ($L_{\rm peak}\sim 3\times 10^{43}$ erg s$^{-1}$) and the relatively nearby redshift, though Ansky is neither the closest nor the most luminous QPE. \cite{Hernandez25} reported the first detection of a highly significant absorption-like spectral feature evolving rapidly within individual bursts, never seen before in a QPE. In particular, recent work on orbiter-disk collisions suggest that the QPE emission may arise from a shock-heated, quasi-spherical debris cloud ejected from the accretion disk, which subsequently undergoes free homologous expansion \citep{Linial23b,Vurm24}. In this manuscript, we analyze the properties of such an expansion, and whether they can effectively reproduce the spectral evolution observed in Ansky. 

In Section~\ref{sec:results}, we present the data products along with phenomenological fits to the spectra. Details of the observations and data analysis procedures can be found in Appendix~\ref{sec:methods}. In Section~\ref{sec:discussion}, we fit collisional ionization and photoionization plasma models to the moderate- and high-resolution spectra, then discuss our interpretation of this feature within the orbiter-disk collision framework. In Section~\ref{sec:conclusion} we make concluding remarks and highlight directions for future study of QPE emission mechanisms.

\begin{figure*}
    \centering
    \includegraphics[width=\textwidth]{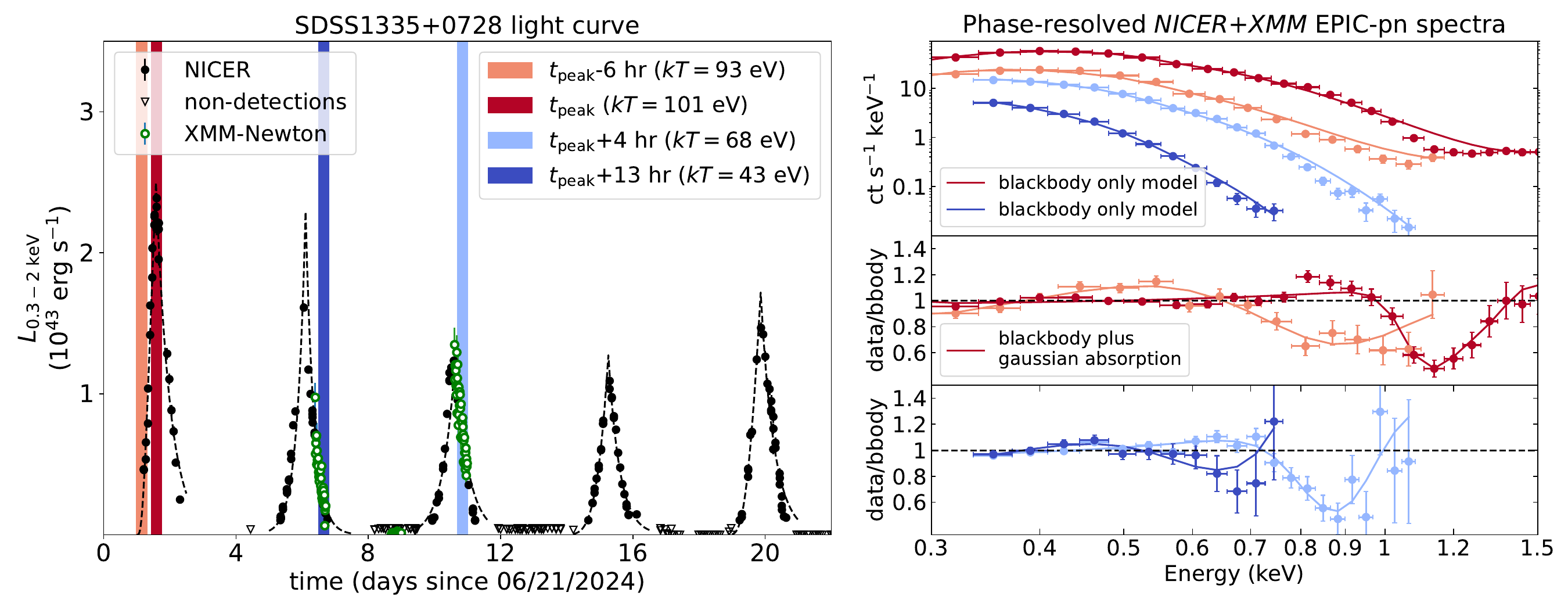}
    \caption{\textbf{Left:} light curve from a subset of the total 60-day monitoring campaign of Ansky. \textbf{Top right:} moderate-resolution spectra at the highlighted phases from \textit{NICER} (red) and \textit{XMM-Newton} EPIC-pn (blue) fit with a blackbody model (solid lines). \textbf{Middle and bottom right:} We separated the \textit{NICER} and \textit{XMM} spectra, and plot the ratios of the data to the best-fitting blackbody model; all spectra show a significant deficit of counts compared to the model between 0.6-1.2 keV. The residuals are well-described by an additional gaussian absorption component (\texttt{gabs}), which are shown by the solid lines.}
    \label{fig:lc_spec}
\end{figure*}

\section{Results} \label{sec:results}

\begin{figure*}
    \centering
    \includegraphics[width=\textwidth]{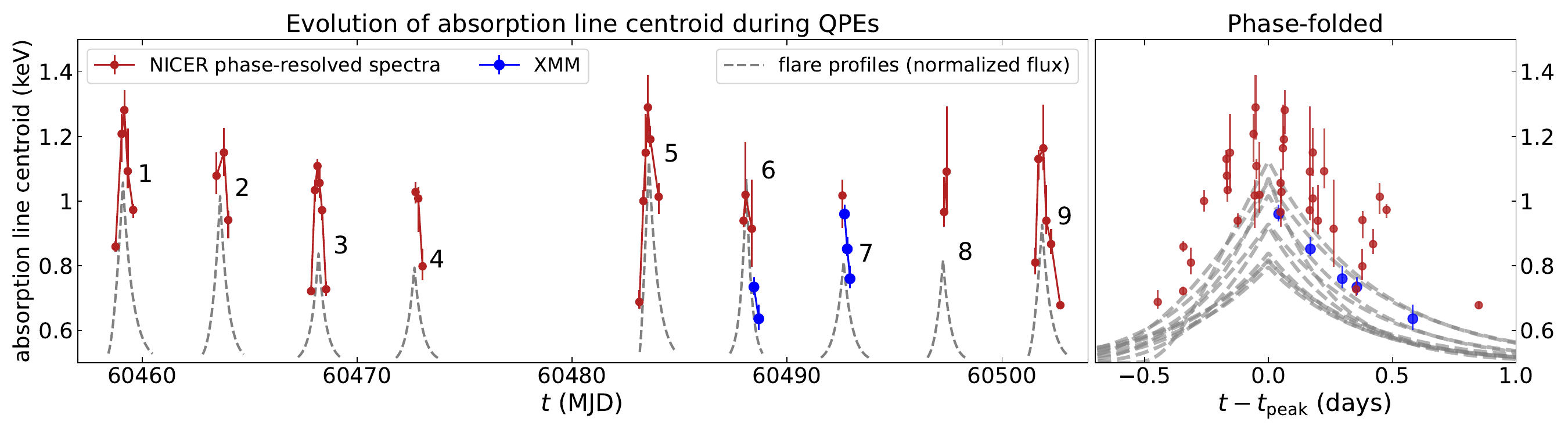}
    \caption{\textbf{Left:} Time evolution of the centroid line energy of the gaussian absorption component. We overplot the flare profiles for comparison. All NICER spectra used for these measurements are shown in Fig.~\ref{fig:ratio_plots}. \textbf{Right:} ``phase-folded'' version of the left panel, i.e. with $t-t_{\rm peak}$ on the $x$-axis. The flare profiles are also normalized to peak flux.}
    \label{fig:evol}
\end{figure*}

\begin{figure*}
    \centering
    \includegraphics[width=\textwidth]{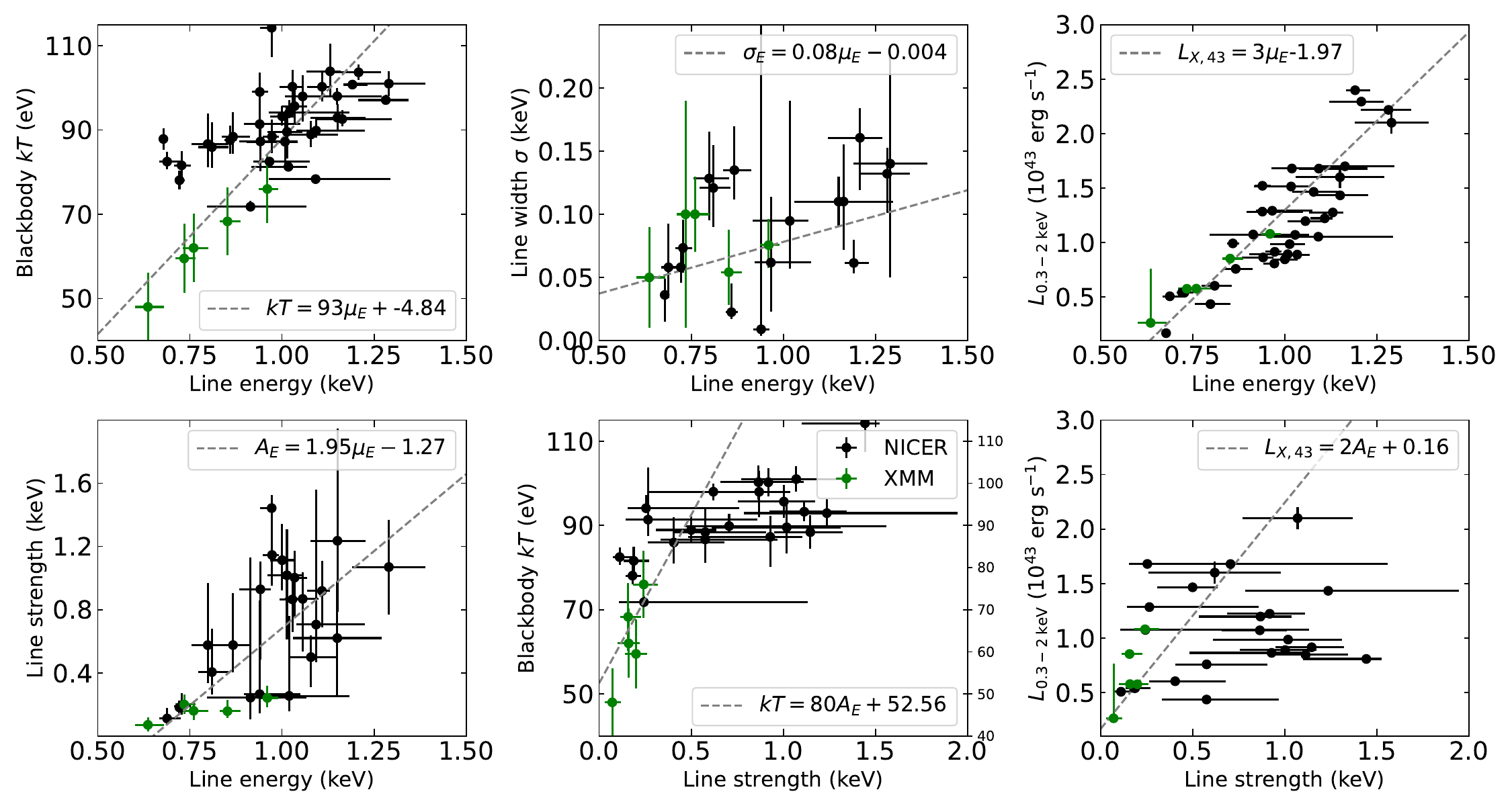}
    \caption{Correlations of the blackbody temperature ($kT$), luminosity ($L_X$), and various parameters of gaussian absorption component: centroid ($\mu_E$), width ($\sigma$), and strength ($A_E$). We note that the \texttt{gabs} model in \texttt{XSPEC} defines the line strength such that the optical depth at the line center is $\tau = A_E/\sigma/\sqrt{2\pi}$. In each plot, we only include the points for which the error calculation converged for both relevant parameters. In the legends we include the best linear fits.}
    \label{fig:corr_plots}
\end{figure*}

QPEs predominantly show blackbody-like X-ray spectra with characteristic temperatures of 50-250 eV \citep{Miniutti19,Giustini20,Arcodia21,Chakraborty21}. They also have a quiescence component well-described by a disk blackbody ($T_{\rm in}\sim 50$ eV) which does not vary as dramatically, suggesting an underlying compact accretion flow. The only observation taken during quiescence with sufficient sensitivity for a detection was a 25.1~ks \textit{XMM-Newton} exposure (OBSID 935191501); fitting with a disk blackbody model, we estimate a quiescence flux of $F_X=2\times 10^{-14}$ erg cm$^{-2}$ s$^{-1}$ ($L_X=3\times 10^{40}$ erg s$^{-1}$), which is a factor of $\sim 500$ fainter than the typical QPE peak \citep{Hernandez25}. Due to the extreme peak-to-quiescence amplitude in this source, we neglected the disk blackbody component for our fits, instead just using a blackbody while accounting for Galactic ISM absorption with a neutral column density fixed to $N_{\rm H}=2.6\times 10^{20}\;{\rm cm}^{-2}$ \citep{HI4PI16}. The faint quiescence and short exposure limits us from searching for discrete spectral features between the eruptions, and the remainder of our analysis focuses on data taken during the eruptions. We performed spectral fitting in \texttt{XSPEC} \citep{Arnaud96}, using a model of \texttt{tbabs}$\times$\texttt{zbbody} (with the blackbody redshifted to the host, $z=0.024$).

The resulting fits show significant absorption-like residuals in the extreme Wien tail, spanning characteristic energies of $\sim 0.6-1.2$ keV with a strong dependence on QPE phase, as reported in \cite{Hernandez25}. We show four sample spectra from different QPEs and different phases in Fig.~\ref{fig:lc_spec}, where the data-to-model ratios being systematically lower than one illustrates the inadequacy of a blackbody-only model. The residuals are seen in \textit{NICER}, \textit{XMM} EPIC-pn, MOS1, MOS2, and RGS spectra, confirming their astrophysical origin. The Wien tail residuals appear even for more complicated choices of continuum model: we tested bulk motion Comptonization (\texttt{bmc}), a disk blackbody (\texttt{diskbb}), and two-component continua comprising a disk plus a blackbody, a double-blackbody, and a blackbody plus Comptonization. None of these choices are able to explain the high-energy residuals.

To explore the time evolution of this feature, we divided the data into eight time bins per QPE defined by 0-25/25-50/50-75/75-100\% of peak flux and vice versa for the decline (see Appendix~\ref{subsec:nicer} for details), then performed phase-resolved fits of the absorption feature with a gaussian profile. We found that a single gaussian absorption line is almost always sufficient to account for the residuals in moderate-resolution spectra, while a gaussian emission line cannot describe the residuals well. We show ratio plots (data divided by blackbody model) from all epochs in which the data quality was sufficient to detect or exclude the presence of this feature in Appendix~\ref{subsec:suppl}. The median 0.3--2 keV C-stat/d.o.f. is 1.54 without the gaussian absorption component, and 1.09 with it. For the median degrees of freedom 104 without the gaussian line, this corresponds to a median fit improvement of $\Delta$C-stat$=$50 for three additional parameters (line centroid, strength, and width). In a handful of spectra, there is an excess emission component red-ward of the gaussian absorption line akin to a P Cygni profile (e.g. QPE 5 peak in  Fig.~\ref{fig:lc_spec}), but generally it is not significant enough to appreciably alter the fit statistic.

With phenomenological fits in hand, in Fig.~\ref{fig:evol} we show the time-evolution of the gaussian centroid energy across QPEs. A recurring overall trend is visible: typically the line energy starts at $\sim 0.7-0.8$ keV early in the rise phase of the QPEs, then increases to a maximum of $\sim 1-1.2$ keV coinciding with peak flux, then finally decreases during the QPE decline phase back down to $\sim 0.6-0.7$ keV. The left panel shows that the higher-luminosity QPEs tentatively coincide with higher-energy lines, though we do not have the data quality to robustly confirm this claim. We do not show the same time-evolution plots for the line strengths and widths, because they are far more poorly constrained due to degeneracies with the background models and low total counts. In general however, they also show a similar time-evolution as the centroid energy, and the blackbody temperature and luminosity are clearly correlated with the line centroid, strength, and width. In Fig.~\ref{fig:corr_plots}, we show the empirical relations between various parameters of the gaussian absorption model evolving as functions of time, as well as linear fits for each pair of variables.

We note that our analysis is not sensitive to features potentially appearing at lower fluxes or during quiescence, because fainter bins have insufficient counts to confirm the presence (or absence) of the gaussian lines. We thus do not comment on whether the absorption disappears altogether at low fluxes, only focusing on the detections at higher $L$.

\section{Discussion} \label{sec:discussion}

\subsection{Physical modeling with ionized absorption/emission} \label{subsec:abs}

\begin{figure}
    \centering
    \includegraphics[width=\linewidth]{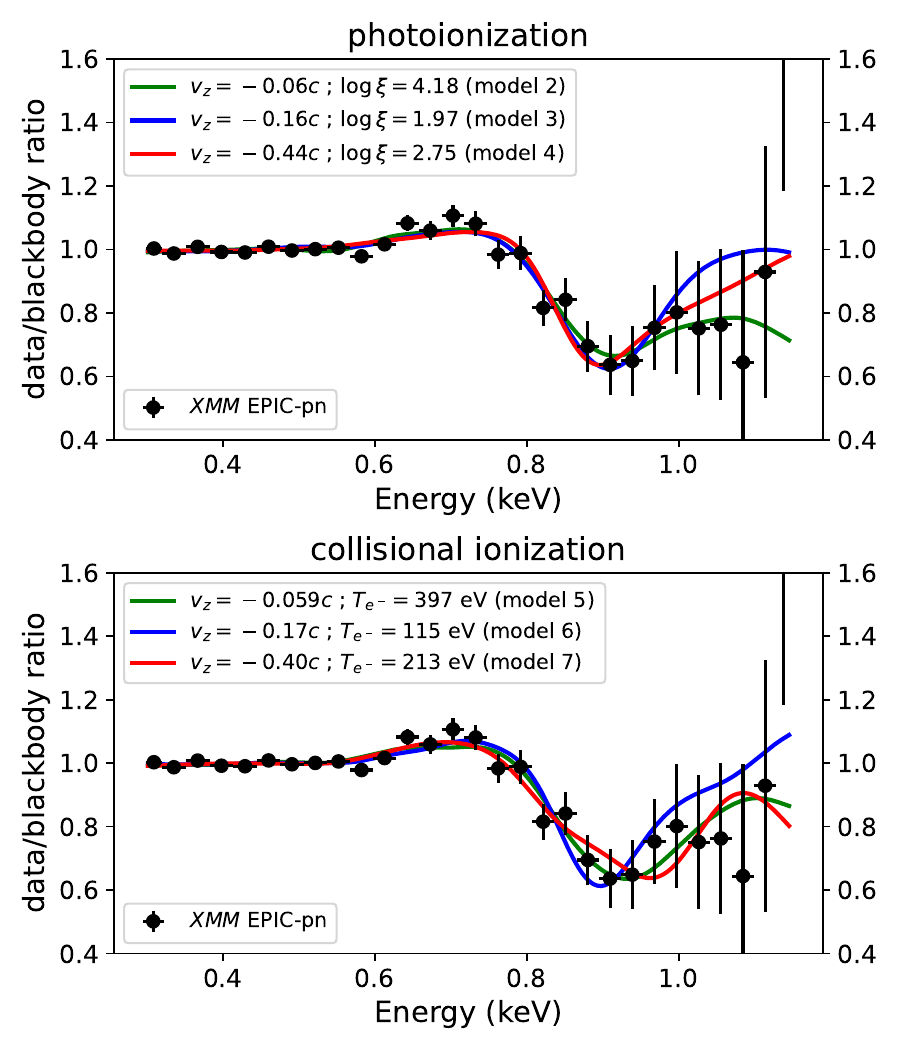}
    \caption{Data-to-blackbody ratios for the \textit{XMM} EPIC-pn spectrum of OBSID 0935191601 (QPE \#7 decline), and six different physical model fits of dense, ionized, blueshifted absorbing plasma (Table~\ref{tab:pn_fits}). All models are statistically near-identical, despite traversing a huge range in bulk velocity ($0.06\lesssim v/c \lesssim 0.4$), broadening ($0\lesssim v_{\rm rms}\lesssim .058c$), and ionization parameter/electron temperature.}
    \label{fig:pn_abs}
\end{figure}

\begin{table*}[ht!]
\centering
\caption{\textit{XMM} EPIC-pn spectral fitting results with photoionization and collisional ionization equilibrium plasma in \texttt{SPEX}. The baseline model (1) is a redshifted blackbody with Galactic ISM absorption. Models (2-4) add photoionized blueshifted plasma absorption (\texttt{HOT}$\times$\texttt{REDS}$\times$\texttt{PION}$\times$\texttt{BB}), while models (5-7) assume collisional ionization equilibrium (\texttt{HOT}$\times$\texttt{REDS}$\times$\texttt{HOT}$\times$\texttt{BB}). We found three statistically comparable solution classes with low (2 \& 5), medium (3 \& 6), and high (4 \& 7) blueshifts.
\label{tab:pn_fits}}
\begin{tabular}{lcccccc}
\toprule
\toprule
& Model & Parameter & Value & Unit & $\pm 1\sigma$ Error & C-stat/d.o.f. \\
\midrule
 \multirow{2}{*}{(1)} & \multirow{2}{*}{Blackbody} & Temperature ($T_{\text{BB}}$) & 67.4 & eV & +0.17/-0.17 & \multirow{2}{*}{247/168} \\
 & & Area ($4\pi R_{\rm BB}^2$) & $1.96\times 10^{24}$ & cm$^2$ & (+3.5/-3.5)$\times 10^{22}$ & \\
\midrule
 \multirow{6}{*}{(2)} & \multirow{2}{*}{Blackbody} & Temperature ($T_{\rm BB}$) & 70.0 & eV & +0.23/-0.30 & \multirow{6}{*}{176/164} \\
 & & Area ($4\pi R_{\rm BB}^2$) & $1.73 \times 10^{24}$ & cm$^{2}$ & (+6.4/-5.9)$\times 10^{22}$ & \\
 & \multirow{4}{*}{\shortstack{Photoionized plasma\\ absorption (\texttt{PION}):\\low-$v_z$ solution}} & Column density ($N_{\rm H}$) & $6.04\times 10^{22}$ & cm$^{-2}$ & (+4.0/-1.4)$\times 10^{20}$ & \\
 & & Ionization ($\log\xi$) & 4.18 & erg s$^{-1}$ cm & +0.24/-0.16 &  \\
 & & RMS velocity ($\sigma$) & 331 & km s$^{-1}$ & +115/-107 & \\
 & & Bulk velocity ($v_z$) & $-1.48\times 10^4$ & km s$^{-1}$ & (+4.8/-3.9)$\times 10^3$ & \\
\midrule
 \multirow{6}{*}{(3)} & \multirow{2}{*}{Blackbody} & Temperature ($T_{\rm BB}$) & 68.9 & eV & +0.23/-0.30 & \multirow{6}{*}{177/164} \\
 & & Area ($4\pi R_{\rm BB}^2$) & $1.96 \times 10^{24}$ & cm$^{2}$ & (+0.70/-1.0)$\times 10^{23}$ & \\
 & \multirow{4}{*}{\shortstack{Photoionized plasma\\ absorption (\texttt{PION}):\\medium-$v_z$ solution}} & Column density ($N_{\rm H}$) & $4.01\times 10^{21}$ & cm$^{-2}$ & (+5.8/-5.3)$\times 10^{20}$ & \\
 & & Ionization ($\log\xi$) & 1.97 & erg s$^{-1}$ cm & +0.31/-0.17 &  \\
 & & RMS velocity ($\sigma$) & 121 & km s$^{-1}$ & +77/-38 & \\
 & & Bulk velocity ($v_z$) & $-4.70\times 10^4$ & km s$^{-1}$ & (+7.8/-0.81)$\times 10^3$ & \\
\midrule
 \multirow{6}{*}{(4)} & \multirow{2}{*}{Blackbody} & Temperature ($T_{\rm BB}$) & 69.1 & eV & +0.61/-0.41 & \multirow{6}{*}{175/164} \\
 & & Area ($4\pi R_{\rm BB}^2$) & $1.82 \times 10^{24}$ & cm$^{2}$ & (+3.9/-4.4)$\times 10^{22}$ & \\
 & \multirow{4}{*}{\shortstack{Photoionized plasma\\ absorption (\texttt{PION}):\\high-$v_z$ solution}} & Column density ($N_{\rm H}$) & $1.58\times 10^{21}$ & cm$^{-2}$ & (+3.9/-2.4)$\times 10^{20}$ & \\
 & & Ionization ($\log\xi$) & 2.75 & erg s$^{-1}$ cm & +0.37/-0.23 &  \\
 & & RMS velocity ($\sigma$) & $1.75\times 10^4$ & km s$^{-1}$ & (+5.3/-4.5)$\times 10^{3}$ & \\
 & & Bulk velocity ($v_z$) & $-1.31\times 10^5$ & km s$^{-1}$ & (+2.3/-5.0)$\times 10^3$ & \\
\midrule
 \multirow{6}{*}{(5)} & \multirow{2}{*}{Blackbody} & Temperature ($T_{\rm BB}$) & 69.2 & eV & +0.27/-0.37 & \multirow{6}{*}{177/164} \\
 & & Area ($4\pi R_{\rm BB}^2$) & $1.83 \times 10^{24}$ & cm$^{2}$ & (+4.6/-4.1)$\times 10^{22}$ & \\
 & \multirow{4}{*}{\shortstack{Collisionally ionized plasma\\absorption (\texttt{HOT}):\\low-$v_z$ solution}} & Column density ($N_{\rm H}$) & $3.85\times 10^{22}$ & cm$^{-2}$ & (+1.8/-0.89)$\times 10^{22}$ & \\
 & & Electron temperature ($T_e$) & 397 & eV & +32/-21 &  \\
 & & RMS velocity ($\sigma$) & 187 & km s$^{-1}$ & +167/-77 & \\
 & & Bulk velocity ($v_z$) & $-1.78\times 10^{4}$ & km s$^{-1}$ & +390/-450 & \\
\midrule
 \multirow{6}{*}{(6)} & \multirow{2}{*}{Blackbody} & Temperature ($T_{\rm BB}$) & 69.1 & eV & +0.29/-0.29 & \multirow{6}{*}{175/164} \\
 & & Area ($4\pi R_{\rm BB}^2$) & $2.25 \times 10^{24}$ & cm$^{2}$ & (+1.1/-0.93)$\times 10^{23}$ & \\
 & \multirow{4}{*}{\shortstack{Collisionally ionized plasma\\absorption (\texttt{HOT}):\\medium-$v_z$ solution}} & Column density ($N_{\rm H}$) & $4.41\times 10^{21}$ & cm$^{-2}$ & (+6.6/-6.1)$\times 10^{20}$ & \\
 & & Electron temperature ($T_e$) & 115 & eV & +6.2/-5.4 &  \\
 & & RMS velocity ($\sigma$) & 399 & km s$^{-1}$ & +180/-150 & \\
 & & Bulk velocity ($v_z$) & $-5.07\times 10^4$ & km s$^{-1}$ & (+4.8/-1.0)$\times 10^3$ & \\
\midrule
 \multirow{6}{*}{(7)} & \multirow{2}{*}{Blackbody} & Temperature ($T_{\rm BB}$) & 69.2 & eV & +0.37/-0.44 & \multirow{6}{*}{176/164} \\
 & & Area ($4\pi R_{\rm BB}^2$) & $1.74 \times 10^{24}$ & cm$^{2}$ & (+2.4/-0.55)$\times 10^{23}$ & \\
 & \multirow{4}{*}{\shortstack{Collisionally ionized plasma\\absorption (\texttt{HOT}):\\high-$v_z$ solution}} & Column density ($N_{\rm H}$) & $4.03\times 10^{21}$ & cm$^{-2}$ & (+1.5/-2.5)$\times 10^{20}$ & \\
 & & Electron temperature ($T_e$) & 213 & eV & +23/-109 &  \\
 & & RMS velocity ($\sigma$) & $1.61\times 10^4$ & km s$^{-1}$ & (+5.0/-5.5)$\times 10^3$ & \\
 & & Bulk velocity ($v_z$) & $-1.21\times 10^5$ & km s$^{-1}$ & (+0.76/-1.5)$\times 10^4$ & \\
\bottomrule
\end{tabular}
\end{table*}

\begin{table*}[ht!]
\centering
\caption{\textit{XMM} RGS spectral fitting results using photoionized plasma models. We first tested a blackbody with cosmological redshift and ISM absorption (model 1). We then used a photoionized emission model (\texttt{HOT}$\times$\texttt{REDS}$\times$\texttt{PION}$\times$\texttt{BB}; model 2) to capture the discrete features from $\sim 0.5-0.7$ keV. The RGS spectrum is consistent with the Wien-tail absorption residuals seen in EPIC-pn, but not with sufficient SNR to constrain models, so we do not include it. In Fig.~\ref{fig:rgs_simult} we showed a composite model (not a fit) consisting of the emission-line plasma below, and one of the absorption components of Table~\ref{tab:pn_fits}. \label{tab:rgs_fits}}
\begin{tabular}{lcccccc}
\toprule
\toprule
& Model & Parameter & Value & Unit & $\pm 1\sigma$ Error & C-stat/d.o.f. \\
\midrule
 \multirow{2}{*}{(1)} & \multirow{2}{*}{Blackbody} & Temperature ($T_{\text{BB}}$) & 73.0 & eV & +0.55/-0.50 & \multirow{2}{*}{1688/1408} \\
 & & Area ($4\pi R_{\rm BB}^2$) & $1.05\times 10^{24}$ & cm$^2$ & (+5.4/-5.4)$\times 10^{22}$ & \\
\midrule
 \multirow{7}{*}{(2)} & \multirow{2}{*}{Blackbody} & Temperature ($T_{\rm BB}$) & 70.1 & eV & +1.4/-1.5 & \multirow{7}{*}{1650/1403} \\
 & & Area ($4\pi R_{\rm BB}^2$) & $1.26 \times 10^{24}$ & cm$^{2}$ & (+1.2/-1.1)$\times 10^{23}$ & \\
 & \multirow{4}{*}{\shortstack{Photoionized plasma\\ (\texttt{PION}) emission}} & Column density ($N_{\rm H}$) & $2.44\times 10^{21}$ & cm$^{-2}$ & (+1.2/-1.1)$\times 10^{21}$ & \\
 & & Ionization ($\log\xi$) & 2.85 & erg s$^{-1}$ cm & +0.26/-0.29 &  \\
 & & RMS velocity ($\sigma$) & $1.03\times 10^4$ & km s$^{-1}$ & (+1.9/-2.1)$\times 10^3$ & \\
 & & Bulk velocity ($v_z$) & $-1.17\times 10^4$ & km s$^{-1}$ & (+1.9/-2.6)$\times 10^3$ & \\
 & & N abundance & 21.7 & solar & +18.5/-11.0 & \\
\bottomrule
\end{tabular}
\end{table*}

\begin{figure*}
    \centering
    \includegraphics[width=\textwidth]{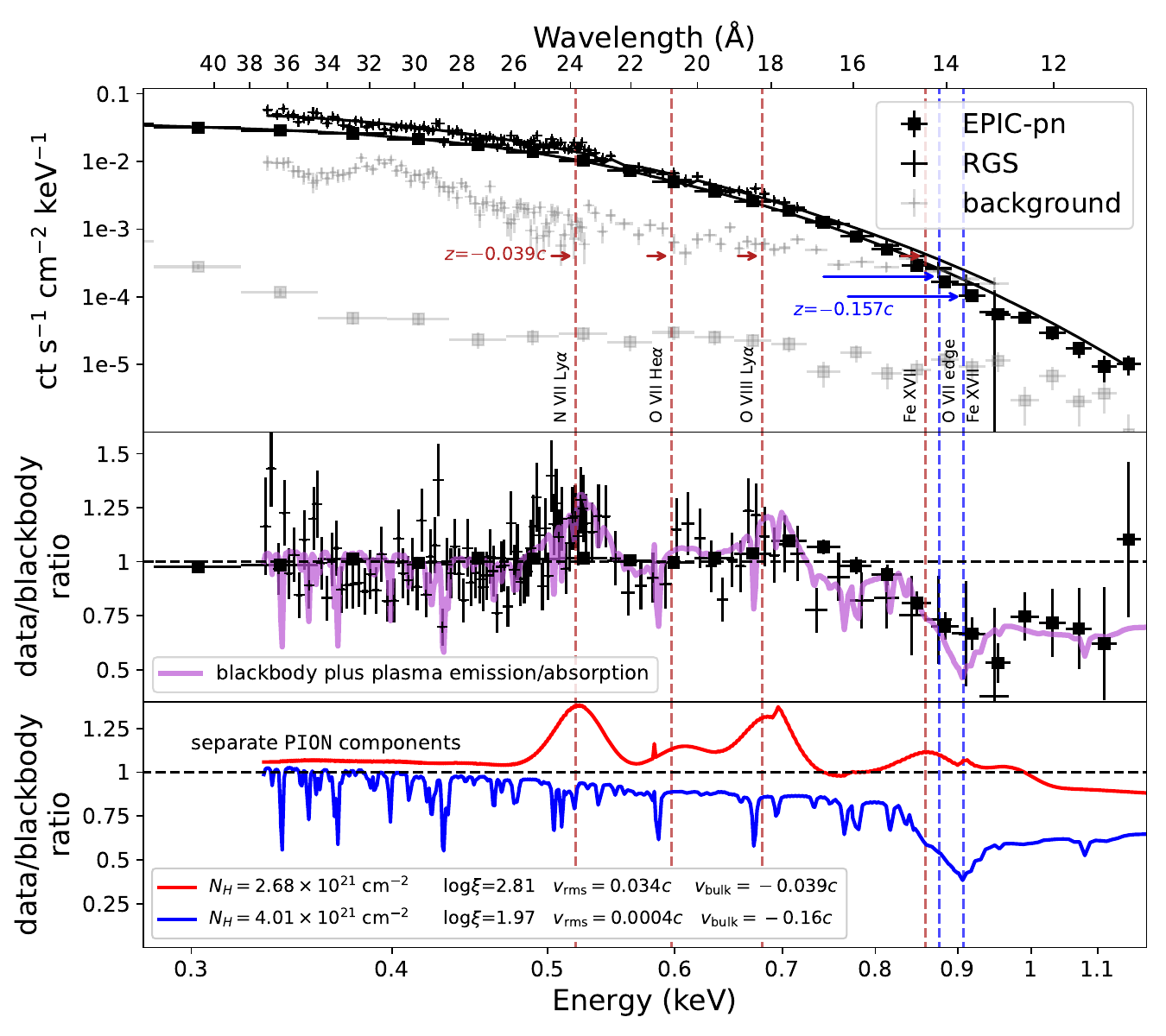}
    \caption{Simultaneous spectra from \textit{XMM} RGS 1+2 (crosses) and EPIC-pn (squares). In the middle panel, we show the ratio of the data to a blackbody-only model. We mark lines and edges of ionized O/N/Fe with the corresponding blueshifts found by model (2) of Table~\ref{tab:rgs_fits}.}
    \label{fig:rgs_simult}
\end{figure*}

Broad features upon a thermal continuum have been seen in the X-ray spectra of nuclear transients before. In the literature, it is common to interpret them as blueshifted ionized emission and/or absorption from a relativistic outflow or accretion disk wind. For instance, similar spectral features have been reported in a handful of tidal disruption events, including ASASSN-14li \citep{Kara18,Ajay24}, 1ES 1927+654 \citep{Ricci21,Masterson22}, AT2022lri \citep{Yao24}, ASASSN-20qc \citep{Kosec23,Pasham24a}, AT2020ksf \citep{Wevers24}, and even the QPE source GSN 069 \citep{Kosec25}. In all of those sources, the line features are seen between $\sim$0.7-1.5 keV. In contrast to Ansky, their properties do not vary rapidly, changing in strength/centroid only over months to years rather than just kiloseconds. It is thus unclear whether they are driven by the same underlying phenomenon, and whether using the same models to describe them is truly appropriate. Nevertheless, to allow comparison with previous results, we performed fits with photoionized and collisionally ionized plasma. Such models are generally best constrained by high-resolution spectra, which we take advantage of here. We performed these fits with the \texttt{SPEX} X-ray spectral fitting code \citep{Kaastra24} due to its built-in capability to perform photoionization calculations self-consistently while varying the illuminating spectrum, using the \texttt{PION} model due to \cite{Mehdipour16}.

We first fit only the EPIC-pn data; the resulting solutions are reported in Table~\ref{tab:pn_fits}. We began by fitting a baseline model of a redshifted blackbody, accounting for Galactic ISM absorption with a neutral plasma (the \texttt{HOT} model with temperature fixed to $10^{-6}$ K) with $N_{\rm H}=2.6\times 10^{20}$ cm$^{-2}$ fixed (a total model of \texttt{HOT}$\times$\texttt{REDS}$\times$\texttt{BB}). We then modeled the Wien-tail absorption features with photoionized (models 2-4) and collisionially ionized (models 5-7) blueshifted absorbing plasmas, i.e. total models of \texttt{HOT}$\times$\texttt{REDS}$\times$\texttt{PION}$\times$\texttt{BB} and \texttt{HOT}$\times$\texttt{REDS}$\times$\texttt{HOT}$\times$\texttt{BB} respectively. We found three solution classes spanning a wide range of blueshifts, around $v\approx -0.06c$ (models 2 and 5), $v\approx -0.16c$ (models 3 and 6) and $v\approx -0.4c$ (models 4 and 7). All absorption models (2-7) result in significant fit improvements of $\Delta$C-stat$=70-72$. It is remarkable that, with the resolution of EPIC-pn, the fits are all statistically near-identical despite the vastly different physical parameters. We plot the change in the model resulting from the addition of absorbing plasma in Fig.~\ref{fig:pn_abs}; the similarity of the models when down-sampled to CCD resolution is manifest.

In Fig.~\ref{fig:rgs_simult}, we show the high-resolution RGS spectrum taken simultaneously with EPIC-pn from ObsID 0935191601, which occurred just after QPE peak and captured most of the decline phase of QPE \#7. The RGS spectrum is consistent with the Wien tail residuals of EPIC-pn, but has neither the signal nor the energy coverage to constrain absorption models at those energies. However, the RGS data also reveal narrower \textit{emission} features at lower energies, further suggesting the presence of blueshifted high-ionization lines of O/N. We therefore proceeded by freezing the absorbing outflow parameters to those found by with EPIC-pn, and adding a lower-velocity emission component in a two-zone plasma model. We used the absorption plasma solution from model 3 of Table~\ref{tab:pn_fits}, because it is the median $v_z$ solution, though we checked that the fit parameters are insensitive to this choice because the absorption features occur predominantly at energies which RGS does not probe. In contrast to the features in EPIC-pn, which are equally well-described by collisional or photoionization, only a photoionized plasma can effectively capture the emission lines seen in RGS. We report the resulting parameters with \texttt{PION} model fits in Table~\ref{tab:rgs_fits}; the addition of the emission-line plasma results in a significant fit improvement of $\Delta$C-stat$=$38. Another noteworthy result from our high-resolution spectroscopy is a large required nitrogen over-abundance of $\sim 21.7\times$ solar. Significant CNO processing is also seen in GSN 069---which is the only other QPE with robust measurements \citep{Sheng21,Kosec25}---as well as in many tidal disruption events from both ultraviolet and X-ray spectroscopy \citep{Cenko16,Kochanek16,Yang17,Miller23}.

Before proceeding, we again stress that our absorption fits are not unique, as the EPIC-pn energy resolution cannot definitively constrain the outflow velocity, density, ionization state, broadening or ionization process. Doing so would require high-resolution spectroscopy, with high enough fluxes/long enough exposures to generate enough signal to capture the absorption features well into the Wien tail. The purpose of our fits is to illustrate the presence of dense ionized plasma with relativistic outflow velocities, as has been seen in the literature before. Our fits do \textit{not} readily provide an explanation for the time-evolution for the features, in particular the rise-and-decay of the outflow velocity one would infer from Fig.~\ref{fig:evol}.

\subsection{Relation to EMRI models} \label{subsec:emri}

In the previous section, we have shown that the broad absorption features in our data can be described equally well by collisionally ionized or photoionized equilibrium plasmas, with a huge possible range of blueshifts $0.06\lesssim v/c \lesssim 0.4$. All models face some difficulties: for example, those with low/medium blueshifts (2-3 and 5-6) require unusually low velocity broadening compared to their bulk velocities, while the models with high blueshifts (4 and 7) require extreme energetics. Without direct high-resolution spectroscopic constraints on the absorption profiles, it is difficult to make further observational progress and exclude or favor a particular solution. We therefore turn to arguments based on theoretical grounds, and discuss the characteristic blueshifts, broadening, and ionization states we expect \textit{may} be produced within the emerging orbiter-disk collision paradigm invoked for QPEs.

We refer the reader to Fig. 1 of \citealt{Vurm24} for a relevant physical picture of the QPE emission source. Each orbiter-collision results in a highly supersonic radiation-mediated shock, in which a cross-section of debris from the accretion disk is quickly raised to higher temperature, pressure, and energy density than the surrounding disk material. This over-pressurized debris breaks out of the disk midplane, proceeding to expand quasi-spherically due to radiation pressure from the photons generated in the early phases following the shock. It is the photon generation and subsequent expansion of this ejected disk material, rather than the star or the disk themselves, which is proposed to power the QPE emission. In this case, the absorption line blueshift is set by the \textit{debris expansion velocity} rather than the bulk velocity of the ejected material. Modeling the time-evolving properties of the ionization features with an accretion disk wind or relativistic outflow, rather than a homologous expansion, would be an interesting exercise, but is outside the scope of this work.

Over the course of the ejecta expansion, the outer layers experience a significant luminosity from the inwardly propagating photon diffusion surface, resulting in significant ionizing flux throughout the entire eruption duration. We thus expect photoionization processes to dominate over collisional ionization, as the latter contributes only during the much briefer orbiter-disk impact duration. However, we emphasize that this is a geometry- and model-dependent assumption. If a spherical homologous expansion is \textit{not} the correct picture for QPEs in Ansky, one could reduce the contribution of photoionization processes, in which case Models (5-7) of Table~\ref{tab:pn_fits} may be the more appropriate ones to consider. Nevertheless, for the remainder of the discussion we focus on interpreting our results within the homologous expansion picture, and defer further considerations of collisional ionization-dominated ejecta to future work.

Previous studies have examined the evolution and energetics of the debris via analytical arguments \citep{Linial23b,Franchini23,linial24c}, as well as hydrodynamical and Monte Carlo simulations \citep{Yao25,Vurm24}. We proceed by briefly describing the key properties of these models, directly considering their implications for spectral line formation, and comparing qualitatively to the features observed in our data.

\subsection{Energetics of the shock breakout and homologous expansion phase} \label{subsec:energetics}

The total energy budget powering each QPE, only a fraction of which escapes as radiation, is set by the dissipated orbital energy during each collision ($E_{\rm tot}$). Initially, this is entirely deposited in the form of internal energy ($E_{\rm int}$), that is, heat plus the energy of trapped radiation. Part of this internal energy is subsequently radiated away as photons powering the observed eruptions ($E_{\rm rad}$), while a larger portion goes into adiabatic losses driving the ejecta expansion (and is thus converted into bulk kinetic energy, $E_{\rm kin}$). The total energy at any given time following the orbiter-disk collision is therefore partitioned among:
\begin{equation*}
    E_{\rm tot} = E_{\rm rad} + E_{\rm kin} + E_{\rm int}.
\end{equation*}
At early times, we have $E_{\rm tot,0} = E_{\rm int,0}$, and as the eruption progresses, we quickly tend towards $E_{\rm int} \ll E_{\rm rad} < E_{\rm kin}$.

Following the shock breakout from the disk, acceleration of the ejecta occurs over just a few radial doubling times \citep{Vurm24}. After this initial acceleration phase, we assume that the debris follows passive homologous expansion, meaning it adopts a quasi-spherical profile which expands self-similarly with $v\propto r/t$. The ejecta start from an initial radius $R_0$ set by the geometrical cross-section of the stellar EMRI \citep{Linial23b} or the Bondi-Hoyle cross-section in the case of a stellar-mass BH \citep{Franchini23}, with an initial thickness $\Delta R_0\sim R_0$. At any given time, the maximum debris radius has the maximum radial velocity, $r/R_{\rm max} = v/v_{\rm max}$, while the minimum radius has $v_{\rm min}\approx 0$, thus remaining at $R_0$ while the outer boundary expands to $R_{\rm max} = v_{\rm max}t$ upon entering the homologous phase. The properties of material undergoing homologous expansion have been studied for decades in several other contexts, from stellar winds \citep{Sobolev60,Castor70} to supernovae \citep{Arnett80,Matzner99,Kasen06}, and we draw upon much of the existing literature for the particular application of QPEs.

Relating $E_{\rm kin}$ to an expansion velocity first requires assuming a density profile. Motivated by the simulations of \cite{Vurm24}, as well as the convention in modeling supernovae \citep{Matzner99}, we adopt a broken power-law form for the ejecta density:
\begin{align}
\rho(r,t) &= \rho_0(t) \begin{cases} 
       \big(\frac{r}{R_{\rm br}(t)}\big)^{-p_1} & R_0 \leq r\leq R_{\rm br}(t) \\
       \big(\frac{r}{R_{\rm br}(t)}\big)^{-p_2} & R_{\rm br}(t) < r \leq R_{\rm max}(t)
   \end{cases}
   \label{eq:rho}
\end{align}
where $p_1$/$p_2$ are power-law indices chosen to describe the radial density fall-off of the ejecta interior/exterior to some break radius $R_{\rm br}(t)$. The prefactor $\rho_0(t)$ is then determined by requiring the full density profile to integrate to a fixed ejecta mass ($M_{\rm ej}$) at all times:
\begin{align}
    \rho_0 &= \frac{M_{\rm ej}}{4\pi}\bigg[ R_{\rm br}^{p_1}\int_{R_0}^{R_{\rm br}} \frac{dr}{r^{p_1-2}} + R_{\rm br}^{p_2}\int_{R_{\rm br}}^{R_{\rm max}} \frac{dr}{r^{p_2-2}}\bigg]^{-1} \notag \\
    &= \frac{M_{\rm ej}}{4\pi}\bigg[\frac{R_0^3R_{\rm br}^{p_1} - R_0^{p_1}R_{\rm br}^3}{R_0^{p_1}(p_1-3)} + \frac{R_{\rm br}^3R_{\rm max}^{p_2} - R_{\rm br}^{p_2}R_{\rm max}^3}{R_{\rm max}^{p_2}(p_2-3)}\bigg]^{-1}
    \label{eq:rho_0}
\end{align}
The total ejecta kinetic energy is thus given by:
\begin{align}
    E_{\rm kin} &= 2\pi\rho_0\bigg[ R_{\rm br}^{p_1}\int_{R_0}^{R_{\rm br}} \frac{v^2 dr}{r^{p_1-2}} + R_{\rm br}^{p_2}\int_{R_{\rm br}}^{R_{\rm max}} \frac{v^2 dr}{r^{p_2-2}}\bigg] \notag \\
    &= 2\pi\rho_0v_{\rm max}^2 \bigg[\frac{R_0^5R_{\rm br}^{p_1} - R_0^{p_1}R_{\rm br}^5}{R_0^{p_1}R_{\rm max}^2(p_1-5)} + \frac{R_{\rm br}^5R_{\rm max}^{p_2}-R_{\rm br}^{p_2}R_{\rm max}^5}{R_{\rm max}^{p_2+2}(p_2-5)}\bigg]
    \label{eq:E_kin}
\end{align}
where in the last line, we have used $v/v_{\rm max}(t)=r/R_{\rm max}(t)$ upon reaching homologous expansion. The maximum velocity at the debris outer boundary can thus be obtained in terms of $p_1$, $p_2$, $R_0$, $R_{\rm br}$, and $R_{\rm max}$.

Another quantity relevant to observations is the radius evolution of the photosphere, i.e. the surface below which the ejecta becomes opaque. Qualitatively, the shock breakout and subsequent expansion will be characterized by a brief initial phase where the photosphere accelerates along with the ejecta, as internal energy is being converted to kinetic. When the conversion has finished and the ejecta reach passive expansion, the photosphere begins to recede to lower normalized radii within the ejecta as the overall density profile becomes diluted. The photosphere radius in the Eddington approximation is obtained from:
\begin{equation}
    \tau_{\rm ph} = \int_{R_{\rm ph}(t)}^{R_{\rm max}(t)} \kappa\rho dr = \frac{2}{3}
\end{equation}
(see Eq. 17 of \citealt{Arnett80} or Eq. 16 of \citealt{Liu18}). For our assumed broken power-law density profile, this equation assumes a step-function form depending on whether or not $R_{\rm ph} < R_{\rm br}$:
\begin{equation}
    \frac{2}{3} = \kappa_{\rm es}\rho_0\begin{cases} 
       \int_{R_{\rm ph}}^{R_{\rm max}}\big(\frac{r}{R_{\rm br}}\big)^{-p_2}dr \qquad\qquad\qquad\;\; R_{\rm ph} \geq R_{\rm br} \\
       \int_{R_{\rm ph}}^{R_{\rm br}}\big(\frac{r}{R_{\rm br}}\big)^{-p_1}dr + \int_{R_{\rm br}}^{R_{\rm max}}\big(\frac{r}{R_{\rm br}}\big)^{-p_2}dr \\ 
       \qquad\qquad\qquad\qquad\qquad\qquad\qquad\; R_{\rm ph} < R_{\rm br}
   \end{cases}
   \label{eq:Rph_1}
\end{equation}
The two cases can be solved for $R_{\rm ph}$:
\begin{equation}
    R_{\rm ph}(t) = \begin{cases} 
       \bigg[\frac{2(p_2-1)R_{\rm br}^{-p_2}}{3\kappa_{\rm es}\rho_0}+R_{\rm max}^{1 -p_1}\bigg]^{\frac{1}{1-p_2}}\qquad R_{\rm ph} \geq R_{\rm br} \\
       \bigg[\frac{1-p_1}{R_{\rm br}^{p_1}}\bigg(\frac{-2}{3\kappa_{\rm es}\rho_0}+\frac{R_{\rm br}}{1-p_1} + \frac{R_{\rm br}^{p_2}R_{\rm max}^{1-p_2}-R_{\rm br}}{1-p_2}\bigg)\bigg]^{\frac{1}{1-p_1}} \\
        \qquad\qquad\qquad\qquad\qquad\qquad\qquad\; R_{\rm ph} < R_{\rm br}
   \end{cases}
   \label{eq:Rph_2}
\end{equation}

The ejecta will also experience an ionizing luminosity from the photosphere over the course of its expansion. The photoionization state of a plasma is often expressed with the ionization parameter, defined as $\xi \equiv \frac{L}{nr^2}$. In our case, since the debris spans several orders of magnitude in $n$ and $r$, we define an ``average'' density-weighted ionization parameter $\bar{\xi}$:
\begin{equation}
    \bar\xi(t) \equiv \frac{\int_{R_{\rm ph}}^{R_{\rm max}}\xi(t,r)n(t,r)dr}{\int_{R_{\rm ph}}^{R_{\rm max}} n(t,r)dr} = \frac{\mu m_p L(t) \int_{R_{\rm ph}}^{R_{\rm max}}r^{-2} dr}{\int_{R_{\rm ph}}^{R_{\rm max}} \rho(t,r) dr}
    \label{eq:logxi_def}
\end{equation}
where $\mu m_p$ is the mean molecular weight and $L(t)$ is the luminosity over time. We choose to weigh by particle density because it directly sets the optical depth as $\tau = n\sigma$, thus regions of highest $n$ contribute most significantly to the absorption profile. For the luminosity, we adopt an exponential rise-and-decay profile going as:
\begin{equation}
    L(t) = \begin{cases}
        L_{\rm peak}\exp\Big(\frac{t-t_{\rm peak}}{\tau_1}\Big) & t \leq t_{\rm peak} \\
        L_{\rm peak}\exp\Big(\frac{-(t-t_{\rm peak})}{\tau_2}\Big) & t > t_{\rm peak}
    \end{cases}
\end{equation}
where $L_{\rm peak}$ is the peak luminosity reached at a time $t_{\rm peak}$, and $\tau_1$/$\tau_2$ are the $e$-folding times for the rise/decay.

\begin{figure*}
    \centering
    \includegraphics[width=\textwidth]{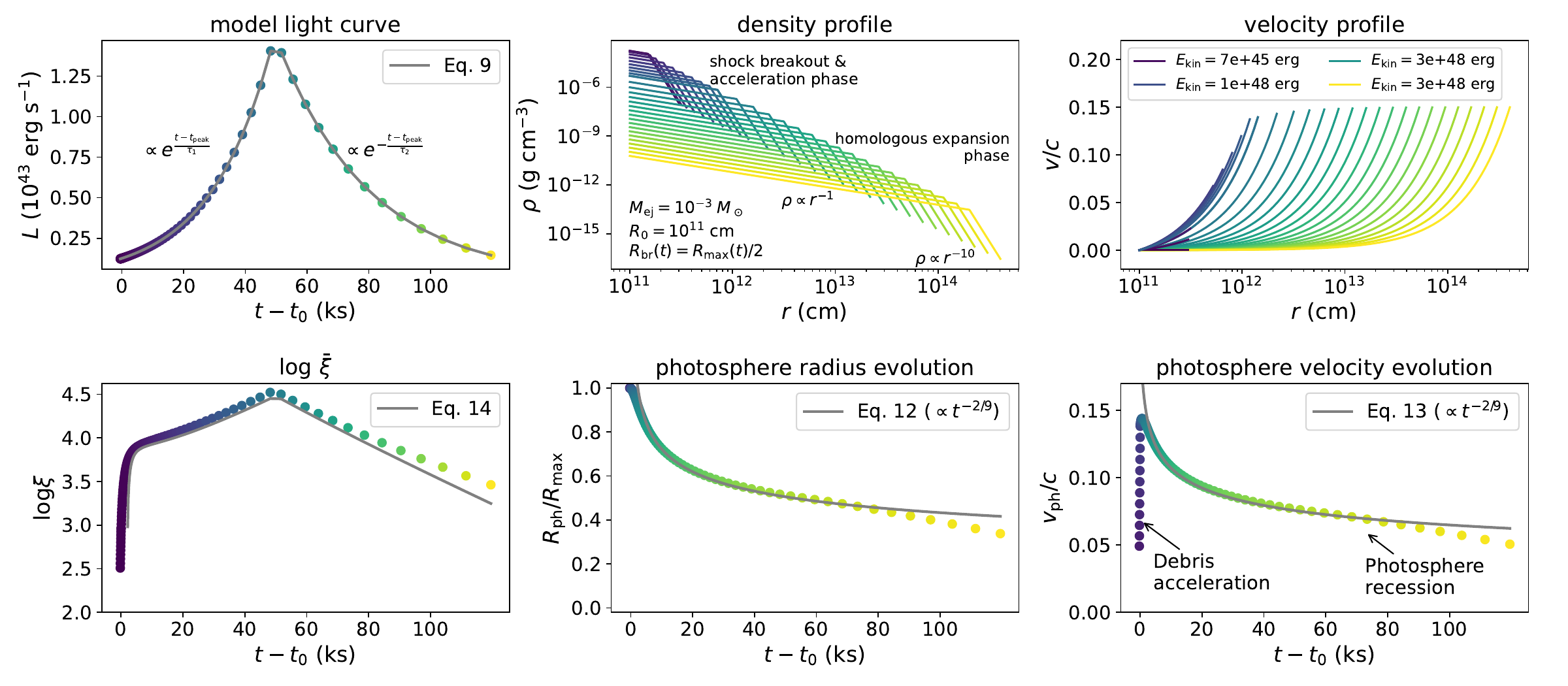}
    \caption{Physical parameters of our analytical toy model for the ejecta evolution. As described in Section ~\ref{subsec:energetics}, we assume the QPE is powered by photons generated by the orbiter-disk collision. We follow the evolution of the escaping overdense ejecta as it is quickly accelerated by radiation pressure, and then reaches passive homologous expansion at a time $t_0$ as light begins to escape (as detailed in \citealt{Vurm24}). The ejecta mass is radially dispersed over the course of the $\sim 10^5$ sec  eruption as the internal energy endowed by the orbiter-disk collision is converted to bulk kinetic energy driving the expansion, converging to $E_{\rm kin}/E_{\rm rad}\sim 3$. Once homologous expansion is reached, the photosphere recedes deeper into the ejecta over time, until the ejecta eventually becomes transparent. The corresponding velocity, $v_{\rm ph}/v_{\rm max}=R_{\rm ph}/R_{\rm max}$, is characterized an initial rise as the ejecta is accelerated, then a decline during the retreat of $R_{\rm ph}$. This two-phase evolution of $\log\bar\xi$ and $v_{\rm ph}$ provide a compelling explanation for the spectral line shifts seen in Fig.~\ref{fig:evol}.}
    \label{fig:ejecta_structure}
\end{figure*}

Having laid out the arguments above, we now use them to construct a toy model for the ejecta evolution over a single QPE of Ansky. We assumed that the debris initially experiences a constant acceleration over a few radial doubling times, then quickly reaches passive expansion for the remaining evolution. For the homologous expansion phase, we adopt fiducial values of $p_1 = 1$, $p_2=10$, $R_0=10^{11}$ cm, $R_{\rm br}(t) = 0.5R_{\rm max}(t)$ (chosen to maintain similarity with Fig. 12 of \citealt{Vurm24}), $M_{\rm ej}=10^{-3}M_\odot$, and $\mu m_p = 0.6m_{\rm H}$. We are pushed to these relatively large ejecta masses to obtain a long enough photon diffusion timescale to reproduce the extreme burst durations of Ansky. Though we adopt $R_0=10^{11}$ cm, our modeling is basically insensitive to any choice of $R_0\lesssim 10^{13}$ cm, as we find expansion velocities reaching $v_{\rm max}\sim 0.15c$, meaning the ejecta will traverse $10^{13}$ cm within just $\sim$2~ks, a small fraction of the typical eruption durations; at this scale, it makes little difference whether the debris started from $R_0=0.1R_\odot$, $R_\odot$ or $10R_\odot$. For our model light curve, we use $L_{\rm peak}=1.5\times 10^{43}$ erg s$^{-1}$, $\tau_1 = 20$ ks, $\tau_2$ = 30 ks, and $t_{\rm peak} = 50$ ks, motivated by the typical profile of the QPEs in Ansky. We plot the evolution of six relevant physical parameters in our toy model---$L(t)$, $\rho(t, r)$, $v(t,r)$, $\bar{\xi}(t)$, $R_{\rm ph}(t)$, and $v_{\rm ph}(t)$---in Fig.~\ref{fig:ejecta_structure}.

For our adopted values, Eq.~\ref{eq:rho_0} yields a density:
\begin{align}
    \rho_0(t) &= \frac{M_{\rm ej}}{4\pi}\bigg(\frac{575R_{\rm max}^3(t)}{7168} - \frac{R_0^2R_{\rm max}(t)}{4}\bigg)^{-1} \notag \\
    &\approx 1.9\times10^{-11}\frac{\rm g}{\rm cm^3} \bigg(\frac{t}{10^4\;\rm s}\bigg)^{-3} \bigg(\frac{v_{\rm max}}{0.15c}\bigg)^{-3} \bigg(\frac{M_{\rm ej}}{10^{-3}M_\odot}\bigg)
    \label{eq:rho_0_1}
\end{align}
where the last line is an approximation which is true by the time the ejecta has reached homologous expansion ($R_0 \ll R_{\rm max}$). We can also write an expression for the velocity of the debris outer edge by combining Eqs.~\ref{eq:rho_0}-\ref{eq:E_kin}:
\begin{align}
    v_{\rm max} &= \sqrt{\frac{2E_{\rm kin}}{M_{\rm ej}}}\bigg(\frac{8960R_0^2R_{\rm max}^2 - 2875R_{\rm max}^4}{4480R_0^4 - 497R_{\rm max}^4}\bigg)^{1/2} \notag \\
    &\approx \sqrt{\frac{11.6E_{\rm kin}}{M_{\rm ej}}} \qquad(R_0\ll R_{\rm max})
   \label{eq:vmax_1}
\end{align}

While we cannot directly determine $E_{\rm kin}$, $E_{\rm rad}$ may be used as a proxy, as it is a measurable quantity from the light curve. Assuming isotropic emission, Ansky radiates an average $\sim 10^{48}$ erg integrated per flare, roughly a factor of 10 higher than the shorter-period QPEs \citep{Hernandez25}. \cite{Vurm24} found via Monte Carlo radiative transfer simulations that by peak light, QPE ejecta converge roughly to $E_{\rm kin}/E_{\rm rad}\sim 3$. So, we can express Eq.~\ref{eq:vmax_1} as:
\begin{align}
    v_{\rm max} &\approx 0.15c \bigg(\frac{E_{\rm rad}}{10^{48}\;\rm erg}\bigg)^{1/2} \bigg(\frac{E_{\rm kin}/E_{\rm rad}}{3}\bigg)^{1/2} 
    \bigg(\frac{M_{\rm ej}}{10^{-3}M_\odot}\bigg)^{-1/2}.
    \label{eq:vmax}
\end{align}
We note that in the left panel of Fig.~\ref{fig:evol}, we tentatively see that higher-luminosity eruptions are indeed associated with higher-energy line centroids, suggesting that $v_{\rm max}$ indeed scales with $E_{\rm rad}$. The $v_{\rm max}$ we infer from Eq.~\ref{eq:vmax} and Table~\ref{tab:pn_fits} imply significantly larger debris velocities than that inferred from the change in $R_{\rm bb}(t)$ as measured from fitting the X-ray spectra, which imply an expansion of only $\sim 1.5\times$ over the course of the eruption \citep{Hernandez25} This discrepancy, however, has been noted before in QPEs, as the inefficiency of photon production and the presence of an electron-scattering atmosphere result in the measured $R_{\rm bb}$ significantly underestimating any true physical length scale or expansion velocity within the debris (for details, see \citealt{Vurm24}).

Conveniently, our choice of parameters results in $R_{\rm ph} > R_{\rm br}$ for most of the burst duration. Eq.~\ref{eq:Rph_2} therefore gives:
\begin{align}
    \frac{R_{\rm ph}}{R_{\rm max}}(t) &= \Big(\frac{\kappa_{\rm es}\rho_0 R_{\rm max}(t)}{6144+\kappa_{\rm es}\rho_0R_{\rm max}(t)}\Big)^{1/9} \notag \\
    &\approx 0.7 \times \bigg(\frac{t}{10^4\;\rm s}\bigg)^{-2/9} \bigg(\frac{v_{\rm max}}{0.15c}\bigg)^{-2/9} \bigg(\frac{M_{\rm ej}}{10^{-3}M_\odot}\bigg)^{1/9}
    % \end{cases}
    \label{eq:Rph_3}
\end{align}
Recalling that $R_{\rm ph}/R_{\rm max}=v_{\rm ph}/v_{\rm max}$, we use Eqs.~\ref{eq:rho_0_1} and~\ref{eq:Rph_3} to write the photosphere velocity over time:
\begin{equation}
    v_{\rm ph}(t) \approx 0.11c \times \bigg(\frac{t}{10^4\;\rm s}\bigg)^{-2/9} \bigg(\frac{v_{\rm max}}{0.15c}\bigg)^{7/9} \bigg(\frac{M_{\rm ej}}{10^{-3}M_\odot}\bigg)^{1/9}
    \label{eq:vph}
\end{equation}
Plugging Eq.~\ref{eq:Rph_3} for the photosphere velocity into Eq.~\ref{eq:logxi_def} also allows us to approximate the time-evolution of the density-weighted ionization parameter:
\begin{align}
    \bar\xi(t) &\approx 9.7\times 10^{4}\;\rm erg\;\rm s^{-1}\;\rm cm \notag \\
    &\times\bigg(\frac{L(t)}{10^{43}\;\rm erg}\bigg)\bigg(\frac{\mu m_p}{0.6m_{\rm H}}\bigg)\bigg(\frac{t}{10^4\;\rm s}\bigg)^{-1}\bigg(\frac{v_{\rm max}}{0.15c}\bigg)^{-1} \notag \\
    &\times\bigg[1.43\bigg(\frac{t}{10^4\;\rm s}\bigg)^{2/9}\bigg(\frac{v_{\rm max}}{0.15c}\bigg)^{2/9}\bigg(\frac{M_{\rm ej}}{10^{-3}M_\odot}\bigg)^{-1/9} - 1\bigg]
    \label{eq:logxi}
\end{align}
We reiterate that Eqs.~\ref{eq:Rph_3}-\ref{eq:logxi} rely on the convenient fact that $R_{\rm ph}\geq R_{\rm br}$ for nearly the entire time-evolution of our toy model, so that we did not have to treat the lower case of Eq.~\ref{eq:Rph_2}. Perturbations from our chosen parameters will quickly run into models in which the photosphere retreat is faster (e.g. for lower $M_{\rm ej}$), and then  Eqs.~\ref{eq:Rph_3}-\ref{eq:logxi} become poor approximations. We defer more general-case modeling, which can explore full expressions for the late-time evolution and/or relax the assumptions on the assumed density profile, to future work.

We plot numerical integrations of $\log\bar\xi$, $R_{\rm ph}/R_{\rm max}$, and $v_{\rm ph}/c$ following Eq.~\ref{eq:Rph_1} across 100 timesteps, along with the analytic expressions provided by Eqs.~\ref{eq:Rph_3}-\ref{eq:vph}, in the bottom middle and right panels of Fig.~\ref{fig:ejecta_structure}. In our toy model, the two-phase evolution of the line profiles in Fig.~\ref{fig:evol} arises due to a combination of the rise and decline in $\log\bar\xi$, changing the relevant line species and ionization balance contributing to the ionization features, while $v_{\rm ph}$ simultaneously decreases, reducing over time the blueshifts at which they appear.

We comment that our toy model outlined above is pushed to somewhat extreme values for the ejecta mass. Maintaining a particle density large enough to have $\log\bar\xi \lesssim 4$ requires $M_{\rm ej}\sim 10^{-3}M_\odot$, in contrast to earlier studies which typically assume $M_{\rm ej}\sim 10^{-5}-10^{-4}M_\odot$ \citep{Linial23b,Vurm24}. Such low masses would result in  $\log\xi \gg 5$, at which point all ions relevant for soft X-ray transitions will be fully stripped of electrons and contribute no absorption. The underlying reason is the extremely long burst durations of Ansky: with such a large $r\sim vt$, the characteristic $\xi=\frac{L}{nr^2} \propto Lr$ (Eq.~\ref{eq:rho_0_1}) is pushed to very high values due to the linear distance scaling. This large radial expansion of the photosphere, by $\sim 50\times$ over the entire eruption duration, also implies there should be significant adiabatic cooling of the ejecta---by a corresponding $0.02\times$---whereas the blackbody temperatures only appear to decrease by a factor $\sim 0.5\times$ \citep{Hernandez25}. It is unclear how this $25\times$ deviation from adiabatic cooling is achieved. One possibility is very inefficient photon production throughout the entire burst, which would result in fewer photons sharing the same (decreasing) internal energy and cause the emission temperature to appear substantially higher than in matter-radiation equilibrium \citep{Linial23a,Vurm24}. However, our toy model requires a more dramatic inefficiency than invoked in prior studies, again owing to the unusually large expansion radii. On its face, solving this would require $v_{\rm max}$ smaller by $\sim 25\times$; however, such a small velocity would be totally incapable of explaining the broad emission lines seen in our data.

With a characteristic $\sim 4\times 10^{48}$ erg released per burst, each QPE dissipates a nontrivial fraction of the total orbital energy available to the EMRI, $E_{\rm orb}=GM_{\rm BH}m/(2a)$. A 1$M_\odot$ EMRI has $E_{\rm orb} \sim1.6\times 10^{51}$ erg, i.e. within 400 eruptions ($\sim 2000$ days) the entire orbital energy would be depleted as the system is promptly driven to merger. A higher-mass EMRI companion would resolve the immediacy of this issue, as with $E_{\rm orb} \gtrsim 10^{53}$ erg the orbit possesses enough energy to remain stable for $>$thousands of orbits. In any case, the dissipation of orbital energy in powering the QPEs appears to be much faster than the gravitational wave timescale; the QPEs in Ansky were first detected in early 2024, so we may expect to see significant evolution in the QPE properties and timescales within just a couple of years. Observations will quickly weigh on this prediction.

We also draw attention to the fact that the large ejecta masses result in a much longer-lived photosphere retreat than in typical QPE models. In the bottom-middle panel of Fig.~\ref{fig:ejecta_structure}, we see that by the end of the 120~ks eruption duration, $R_{\rm ph}\sim 0.4R_{\rm max}$, i.e. a 40\% of the debris has yet to become transparent. We suggest that the further retreat of the photosphere may be accompanied by photons experiencing such large adiabatic losses that they are observed in ultraviolet rather than X-rays, a possibility also suggested in \cite{Linial24b} and \cite{Vurm24}. Remarkably, Ansky shows hints of rapid UV variability unusual for a QPE \citep{Sanchez24,Hernandez25}, though not with sufficiently high cadence or significance to resolve any relation to the X-ray bursts. Follow-up higher-cadence UV monitoring has the potential to reveal whether the X-ray QPEs in this source are indeed accompanied by a longer-lived UV tail.

\subsection{Line formation in expanding QPE atmospheres} \label{subsec:line_form}

Existing models for absorption by a photoionized/collisionally ionized plasma, such as the \texttt{PION} and \texttt{HOT} models used in Section~\ref{subsec:abs} (but also the commonly-used \texttt{XSTAR} model within \texttt{XSPEC}) are designed to compute the transmission and absorption through a \textit{uniform slab geometry} of ionized plasma. This is not appropriate for a spherical homologous expansion experiencing a significant density gradient of several orders of magnitude. Therefore, keeping with our assumptions for the debris density profile and geometry, we turn to the problem of spectral line formation in the expanding spherical QPE atmospheres. Line formation in spherically expanding media has been explored in detail by \cite{Sobolev60}, \cite{Castor70}, \cite{Kasen06}, and others. We describe the effect of the rapidly expanding debris cloud on a simple delta-function line profile as a means of illustrating the qualitative expectation. For much of the following argument, we follow \cite{Castor70}, albeit with some key differences with respect to the line resonance condition and profile.

\begin{figure*}
    \centering
    \includegraphics[width=\textwidth]{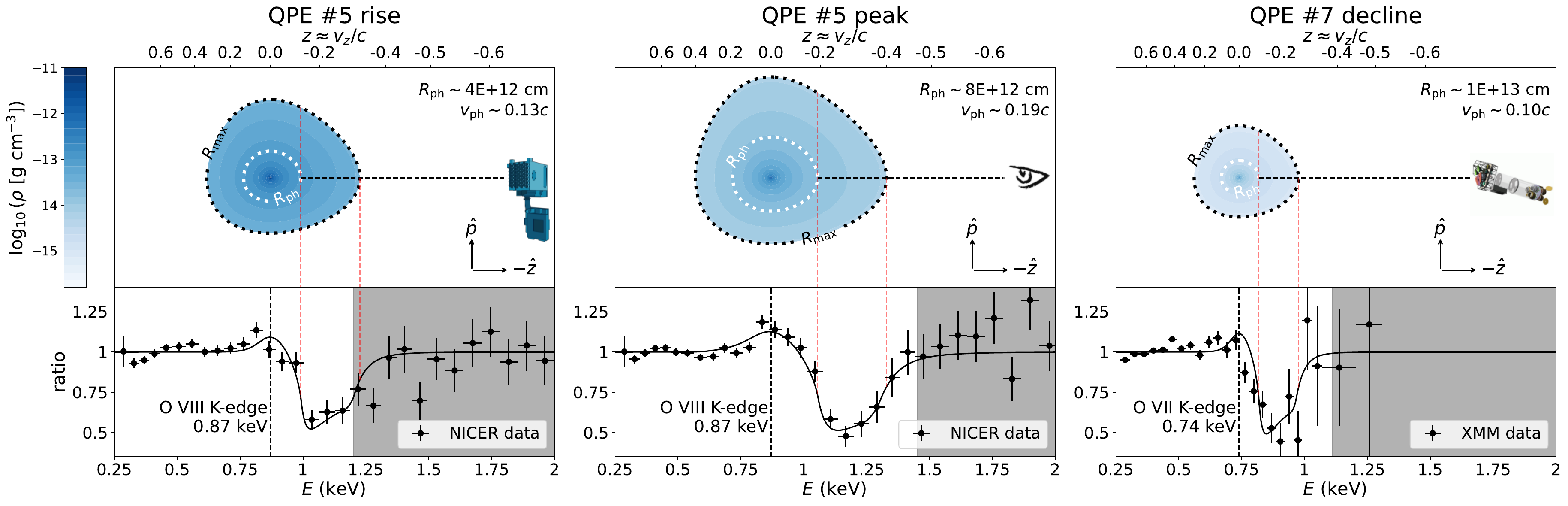}
    \caption{\textbf{Top panels:} schematic of spherically symmetric debris in homologous expansion (the apparent deviation from spherical geometry is due to the nonlinear axes, which correspond to the velocity blueshift at each $z$-coordinate). We indicate contours of both density and radial coordinate $r$, and mark the $z$-coordinate corresponding to the line-of-sight velocity on the top axis. The observer sits at $z=-\infty$, with a sight-line peering through the debris until it becomes occluded by the photosphere at $R_{\rm ph}$. \textbf{Bottom panels:} line profiles corresponding to each choice of debris parameters, along with observational data from \textit{NICER}/\textit{XMM}. They are not fits to the data, and are overplotted only to show the qualitative agreement with the model.}
    \label{fig:schematic}
\end{figure*}

Relying on the assumption of spherical symmetry, we work in a two-dimensional polar coordinate system ($p$, $z$) centered on the ejecta, which is expanding radially in all directions (independent of polar/azimuthal angle). We orient the observer along the $z$-axis at $-\infty$, so that we only consider light rays traveling along $-\hat{z}$. The coordinate axes are shown in the top panels of Fig.~\ref{fig:schematic}; this definition of $z$ coincides conveniently with the observed velocity blueshift, owing to the homologous expansion assumption ($r\propto v_r$, $z\propto v_z$). The cylindrical coordinates ($p$, $z$) are related to the usual polar coordinates ($r$, $\theta$) by $r^2 = p^2+z^2$ and $z=-r\cos\theta$. The radiative transfer equation in this coordinate system reads:
\begin{align}
    \frac{dI(z,p,\nu)}{dz} &= \rho(r)\kappa(\nu,r) [-S(r) + I(\nu, p, z)] \notag \\
    &= n(r)\sigma(\nu,r) [-B_\nu(T(r)) + I(\nu, p, z)] \label{eq:radiative_transfer}
\end{align}
where we have assumed the source function in the atmosphere is an isotropic blackbody spectrum depending only on the radial coordinate within the ejecta. To solve this equation for the intensity $I(\nu,p,z)$, we divide the debris into a discrete boundary at the photosphere, and assume the photosphere radiates a single-temperature blackbody intensity $I_{\rm ph}=B_\nu(T_{\rm ph})$ independent of spatial coordinate within the boundary $R_{\rm ph}$. The solution to the radiative transfer equation in the expanding envelope ($r>R_{\rm ph}$) is thus divided into the contributions from the near and far sides of the photosphere:
\begin{align}
    I(\nu, p, z) &= \int_{n\sigma z}^\infty S\big(\sqrt{p^2+z'^2}\big)e^{\tau(\nu,p,z)-\tau'(\nu,p,z')}d\tau'(\nu,p,z') \notag \\
    &\qquad\qquad\qquad\qquad\mathrm{if}\;p > R_{\rm ph}\ \mathrm{or}\ z > 0 \notag \\
    I(\nu, p, z) &= \int_{n\sigma z}^{-n\sigma\sqrt{R_{\rm ph}^2-p^2}} S\big(\sqrt{p^2+z'^2}\big)e^{\tau(\nu,p,z)-\tau'(\nu,p,z')}d\tau' \notag \\
    &\qquad\qquad + B_\nu(T_{\rm ph})\exp\Bigg(-\int_z^{-\sqrt{R_{\rm ph}^2-p^2}}n\sigma dz'\Bigg) \notag \\
    &\qquad\qquad\qquad\qquad\mathrm{if}\;p < R_{\rm ph}\ \mathrm{and}\ z < 0
\end{align}

To write an expression for $\tau$, we consider a single line with a rest-frame frequency $\nu_0$, and take the limit that its intrinsic width due to thermal/pressure/natural broadening is much smaller than the Doppler broadening induced by rapid expansion of the envelope. Under this approximation, photons emitted along $-\hat{z}$ at a particular energy $\nu$ will experience no line opacity until they come into contact with the particular region moving with $v_z$ satisfying:
\begin{equation}
    1+z = \frac{\nu_0}{\nu} = \bigg(\frac{1+v_z/c}{1-v_z/c}\bigg)^{1/2}.
\end{equation}
We cannot make the usual approximation $1+z\approx v_z/c$, because the ejecta velocities approach appreciable fractions of $c$. This resonance condition requires:
\begin{equation}
    \frac{v(r)\cos\theta}{c} = \frac{v_{\rm max}\sqrt{p^2+z^2}}{cR_{\rm max}}\frac{z}{\sqrt{p^2+z^2}} = \frac{\big(\frac{\nu_0}{\nu}\big)^2-1}{\big(\frac{\nu_0}{\nu}\big)^2+1},
\end{equation}
We refer to the $z$-coordinate satisfying this condition with $z\equiv z_0(p, \nu_0, \nu)$, the corresponding $r_0\equiv \sqrt{p^2+z_0^2}$, and the temperature at this radius $T_0\equiv T(r_0)$. In this narrow-line limit, the optical depth under this condition can be expressed with a delta function:
\begin{align}
    \tau(z,p,\nu) &= \int_z^\infty n(z',p)\sigma\delta\bigg[\frac{v_{\rm max}z'}{cR_{\rm max}} - \frac{\big(\frac{\nu_0}{\nu}\big)^2-1}{\big(\frac{\nu_0}{\nu}\big)^2+1}\bigg] dz' \notag \\
    &= n(r_0)\sigma \frac{cR_{\rm max}}{v_{\rm max}} H(z_0-z)
\end{align}
where $H$ is the Heaviside step function, and $\sigma$ is the \textit{a priori} known cross-section of the line of interest. The values for some relevant transitions, which we use in our modeling, are $\sigma \sim 10^{-19}$ cm$^{-2}$ for the O {\sc viii} 0.87 keV edge, and $\sigma \sim 2.4\times 10^{-19}$ cm$^{-2}$ for the O {\sc vii} 0.74 keV edge \citep{Kaastra08}.

Then, using this form for $\tau$, we can write the intensity at the location of the observer ($z=-\infty$) as:
\begin{align}
    I(\nu, p, -\infty) &= B_\nu(T_0)\Big(1-e^{-n(r_0)\sigma cR_{\rm max}/v_{\rm max}}\Big) \notag \\
    &\qquad\qquad\qquad\mathrm{if}\;p > R_{\rm ph}\ \mathrm{or}\ z > 0 \notag \\
    I(\nu, p, -\infty) &= B_\nu(T_0)\Big(1-e^{-n(r_0)\sigma cR_{\rm max}/v_{\rm max}}\Big) \notag \\
    &\qquad\qquad + B_\nu(T_{\rm ph})e^{-n(r_0)\sigma cR_{\rm max}/v_{\rm max}} \notag \\
    &\qquad\qquad\qquad\mathrm{if}\;p < R_{\rm ph}\ \mathrm{and}\ z < 0
\end{align}

The observed flux resulting from this intensity is the integral of the above expression over the angular coordinate:
\begin{align}
    F_\nu &= 4\pi \int_0^{R_{\rm ph}} \bigg[B_\nu(T_0)\Big(1-e^{-n(r_0)\sigma cR_{\rm max}/v_{\rm max}}\Big) \notag \\
    &\qquad + B_\nu(T_{\rm ph})e^{-n(r_0)\sigma cR_{\rm max}/v_{\rm max}}\bigg] 2\pi pdp \notag \\
    &\qquad + 4\pi \int_{R_{\rm ph}}^{\infty} B_\nu(T_0)\Big(1-e^{-n(r_0)\sigma cR_{\rm max}/v_{\rm max}}\Big) 2\pi pdp
    \label{eq:line_profile}
\end{align}
where $n(r)$ is computed from Eqs.~\ref{eq:rho} and \ref{eq:rho_0_1}, $v_{\rm max}$ from Eq.~\ref{eq:vmax_1}, $R_{\rm ph}$ from Eq.~\ref{eq:Rph_3}, an $T(r)/T(R_{\rm ph}) \propto [\rho(r)/\rho(r_{\rm ph})]^{\gamma-1}$ for adiabatic expansion (where $\gamma = 4/3$ for a radiation-mediated shock). Meanwhile, the photosphere emits a spatially uniform intensity, $F_{\nu,\rm ph} \propto R_{\rm ph}^2 B_\nu(T_{\rm ph})$. Computing a line profile thus amounts to solving Eq.~\ref{eq:line_profile}, and dividing by the flux of a single-temperature blackbody.

We discretized and numerically computed Eq.~\ref{eq:line_profile} on a grid of $\nu,p$ values, and show ratio plots of $F_\nu/F_{\rm \nu,ph}$ in Fig.~\ref{fig:schematic}. For our modeling, we assumed a solar abundance of oxygen ions, and that either the hydrogen-/helium-like state dominates the ionic population based on the energies at which the absorption is observed. In the rise/peak spectra where the residuals appear just above 0.87 keV, we assume the line profile is dominated by the O {\sc viii} K-edge; for the decline spectrum with absorption above 0.74 keV, we assume it is dominated by the O {\sc vii} K-edge. We emphasize that we make explicit assumptions about which line species we are observing, but that our assumptions are consistent with the photoionization modeling in Section~\ref{sec:results}, i.e. that the range of observed blueshifts occupies $\sim 0.1-0.4c$ in agreement with the \texttt{PION} models reported in Table~\ref{tab:pn_fits}. In reality, multiple line species are likely to contribute, resulting in a complicated blend of absorption residuals which likely varies across the ejecta radial coordinate; our modeling is meant only to corroborate the general picture of homologous expansion, and provides good qualitative agreement with the data. The photosphere velocities we obtain ($v_{\rm ph}\sim 0.1-0.2c$) agree reasonably well with Eq.~\ref{eq:vph} (recall also that QPE \#5 is exceptionally bright, and should thus have a higher kinetic energy budget). The $v_{\rm max}\sim 0.2-0.4c$ one would infer from our model is larger than the characteristic value $v_{\rm max}\sim 0.15c$ we assumed in Section~\ref{subsec:energetics}; this is likely due to our approximation of delta-function line width and/or one species contributing, when in reality there will be some additional broadening and multiple blended lines.

The line shapes are quite sensitive to the break radius $R_{\rm br}$ and the exponents $p_1$, $p_2$, as these most directly determine the distribution of the line-of-sight obscuring matter encountered by light rays from the photosphere. However, the profiles do not uniquely determine the underlying physical parameters ($\rho$, $\xi$, $T$) without further constraints on the structure of the expanding envelope. For example, one could achieve the same effective $\tau$ by decreasing ionization while increasing mass density, or increasing mean molecular weight while decreasing mass density, all of which we have simply assumed. A full solution would require self-consistently computing the radiation energy density and temperature structure within each radial shell of the ejecta, then performing radiative transport simulations at each time step to provide the corresponding change in radius and escaping radiation as a function of time. While this is certainly an interesting avenue for further study, and may yield better constraints on the ejecta properties given more data, it is beyond the scope of this paper. Here we have merely aimed to show that the broad, time-varying line profiles seen in QPEs \textit{can} be explained by the homologous expansion phase naturally following an orbiter-disk shock breakout.

Fig.~\ref{fig:schematic} illustrates pictorially that the dominant factor in determining the location of the absorption line centroid is the expansion velocity of the photosphere ($v_{\rm ph}$), as discussed in Section~\ref{subsec:energetics}: only exterior to $R_{\rm ph}$ can we see absorption features, so $v_{\rm ph}$ provides a lower bound to the line centroid at any given time. At the same time, $v_{\rm max}$ provides an upper bound, because it corresponds with the outer edge of the ejecta, exterior to which there is no obscuring material. The true blueshifts corresponding to the phenomenological gaussian line centroid reported in Fig.~\ref{fig:evol} must lie between $v_{\rm ph}$ and $v_{\rm max}$, but making a stronger statement than this is difficult.

\section{Conclusion} \label{sec:conclusion}

We have reported and studied, for the first time, the rapid time evolution of broad absorption lines in the spectra of a QPE (Fig.~\ref{fig:lc_spec}), seen in moderate-resolution spectra with both \textit{NICER} and \textit{XMM} EPIC-pn/MOS1/MOS2. The line centroids shift dramatically over the course of each eruption in a recurring rise-and-decline pattern (Fig.~\ref{fig:evol}), as do the line strength and width. They are all correlated with the continuum temperature and luminosity (Fig.~\ref{fig:corr_plots}). As is common in the literature, we fit the EPIC-pn spectra with physical models of dense, blueshifted, ionized plasma occluding the thermal continuum emission (Fig.~\ref{fig:pn_abs}; Table~\ref{tab:pn_fits}). Although the resolution of CCD spectra are not sufficient to uniquely determine the physical properties of the absorbing medium, and we find six different statistically comparable solutions, we can infer outflow column densities of $N_{\rm H} \gtrsim 10^{21}$ cm$^{-2}$ and blueshift velocities of $0.06\lesssim v/c \lesssim 0.4$. The data can be explained by either a collisionally ionized plasma with electron temperature $115\lesssim T_e\lesssim 397$ eV, or a photoionized plasma with ionization parameter $1.97\lesssim \log\xi \lesssim 4.18$. With high-resolution X-ray spectra from \textit{XMM} RGS, we also detected \textit{emission} lines at lower blueshift, which are well-described by a photoionized plasma with $N_{\rm H}\sim 2.4\times 10^{21}$ cm$^{-2}$, $\log\xi \sim 2.81$, and bulk velocity $v/c\sim -0.039$. The relative line strengths point toward a high nitrogen overabundance of $\sim 21.7\times$ solar (Fig.~\ref{fig:rgs_simult}; Table~\ref{tab:rgs_fits}), as has been seen in other tidal disruption events and the QPE-emitting galaxy GSN 069. Our RGS spectra do not retain enough signal at high enough energies to constrain the absorption features, but future X-ray observations of the QPEs in Ansky offer the enticing possibility of making these constraints.

We proceeded by interpreting the absorption features within the emerging theoretical paradigm that QPEs are powered by orbiter-disk collisions. In this picture, the evolution of the QPE emission and its corresponding spectral evolution can be analyzed with a toy model of shock-heated debris undergoing spherical homologous expansion after breaking out from the accretion disk (e.g. \citealt{Franchini23,Linial23b,Vurm24}). Within the homologous expansion picture, the blueshifts of the absorption lines are linked to the debris \textit{expansion velocity} rather than bulk velocity. We outline an analytical toy model for a single burst of Ansky, with characteristic ejecta masses $M_{\rm ej}\sim 10^{-3}M_\odot$ and expansion velocities up to $0.15c$ at the debris outer edge. The densities attained result in reasonable ionization parameters ($\log\xi\lesssim 4.5$) and explain the rise-and-decline of the of the absorption line centroids (Fig.~\ref{fig:evol}) with the two-phase evolution in ionization parameter coupled with the steady retreat of the photosphere ($v_{\rm ph}\propto t^{-2/9}$) which further accelerates the decline in line energy (Fig.~\ref{fig:ejecta_structure}).

The large ejecta masses required to maintain a low enough $\xi$ for soft X-ray absorption mean the retreat of the photosphere into the ejecta progresses much more slowly than in other QPEs, and within our toy model $R_{\rm ph}$ only reaches about $0.4R_{\rm max}$ within the 120~ks eruption duration (Fig.~\ref{fig:ejecta_structure}). We suggest that the further retreat of the photosphere may be accompanied by photons experiencing large enough adiabatic losses that they are released in ultraviolet, an intriguing possibility suggested in \cite{Linial24b} and \cite{Vurm24}, and hinted by the unusually large and rapid (but unresolved as of yet) UV variability of the source \citep{Sanchez24,Hernandez25}.

Our toy model dissipates $\sim 4 \times 10^{48}$ erg in kinetic plus radiated energy per burst. For comparison, $1 M_\odot$ EMRI at $500R_g$ possesses a total orbital energy of $\sim 1.6\times 10^{51}$ erg, and can thus support  $\lesssim 400$ eruptions of Ansky. In this case, we should observed significant, rapid evolution in the QPE timescales and properties within just a few years. Ongoing \textit{NICER} and \textit{XMM} monitoring will imminently test this hypothesis.

The spherical homologous expansion geometry differs significantly from the slab geometry used for standard ionized plasma models (e.g. \texttt{HOT} and \texttt{PION} in \texttt{SPEX}, \texttt{XSTAR} in \texttt{XSPEC}), making it difficult to directly interpret fits with those models within the orbiter-disk collision paradigm. We therefore addressed the problem of line formation in expanding QPE atmospheres using the Sobolev approximate commonly adopted for stellar winds and supernovae. Based on our toy model assumptions for ejecta mass, kinetic energy, ionization balance, and density profile, we are indeed able to qualitatively reproduce the spectral features of Ansky within a spherical homologous expansion geometry (Fig.~\ref{fig:schematic}), though we emphasize we did not directly fit the data due to the degeneracies and assumptions involved. Moreover, our order-of-magnitude picture faces a significant problem with the temperature evolution of Ansky: the measured blackbody temperatures decrease by $\sim 2\times$ over the course of an eruption, which is significantly less than the adiabatic cooling expected from the $\sim 50\times$ photosphere expansion predicted during the homologous expansion phase (Fig.~\ref{fig:ejecta_structure}). It is unclear how to reconcile this while maintaining the relativistic expansion velocities to explain the observed spectral features. Further work is thus required, and well-motivated by the quickly-growing QPE population, to construct models which self-consistently compute the time-evolving ionization, line opacity, and escaping radiation in QPEs. Here we have constructed a proof-of-concept model which seems able to describe the data well, while naturally extending existing models for QPEs; in future works, we will continue to further study and extend these arguments.

\section*{Acknowledgements}
We thank Brian Metzger, Itai Linial, Indrek Vurm, Daniele Rogantini, and Frits Paerels for useful discussions regarding this work and suggestions on an earlier version of this draft. We also thank Keith Gendreau, Zaven Arzoumanian, and the \textit{NICER} scheduling and operations team for prompt, high-cadence monitoring of the QPEs in Ansky, without which this work would not be possible. We thank the anonymous referee for insightful comments which improved the quality of the paper.
GM acknowledges support by grant PID2020-115325GB-C31 funded by MCIN/AEI/10.13039/50110001103.
RA was supported by NASA through the NASA Hubble Fellowship grant \#HST-HF2-51499.001-A awarded by the Space Telescope Science Institute, which is operated by the Association of Universities for Research in Astronomy, Incorporated, under NASA contract NAS5-26555.
MG is funded by Spanish MICIU/AEI/10.13039/501100011033 grant PID2019-107061GB-C61.
CR acknowledges support from Fondecyt Regular grant 1230345, ANID BASAL project FB210003 and the China-Chile joint research fund.

\appendix
\counterwithin{figure}{section}
\counterwithin{table}{section}

\section{Observations and data reduction} \label{sec:methods}
\subsection{NICER} \label{subsec:nicer} 
The \textit{NICER} X-ray Timing Instrument \citep{Gendreau16} aboard the International Space Station observed Ansky for a total of 278~ks across 60 Target of Opportunity (ToO) observations (ObsIDs 7204490101-7204490160) from May 19-July 20, 2024. We exactly followed the time-resolved spectroscopy approach for reliable estimation of source light curves outlined in Section 2.1 of \cite{Chakraborty24}. The light curve thus generated was presented in \cite{Hernandez25}, and we reproduced part of it in Fig.~\ref{fig:lc_spec}.

We then fit the data with exponential rise-/decay-profiles to determine the QPE phase of each \textit{NICER} snapshot. Eruptions were divided into eight bins, with four bins during the rise corresponding to 0-25/25-50/50-75/75-100\% of peak flux, and vice-versa for the decline. We determined good-time intervals within each observation using the \texttt{nimaketime} routine, then dividing them into eight segments corresponding to each QPE phase bin. In some cases when the \textit{NICER} ObsID transitions in the middle of a phase bin, we used the command \texttt{niobsmerge} to combine data products across observations before extracting spectra. We fit over a large energy range (0.25--10 keV) for data taken in orbit night, and a restricted range (0.38--10 keV) during orbit day, to allow accurate estimation of the background with the \texttt{SCORPEON}\footnote{\href{https://heasarc.gsfc.nasa.gov/docs/nicer/analysis_threads/scorpeon-overview/}{https://heasarc.gsfc.nasa.gov/docs/nicer/\\analysis\_threads/scorpeon-overview}} model. \texttt{SCORPEON} is a semi-empirical, physically motivated background model which explicitly includes components for the cosmic X-ray background as well as non X-ray noise events (e.g. precipitating electrons and cosmic rays) and can be fit along with the source to allow joint estimation of uncertainties. We grouped our spectra with the optimal binning scheme of \cite{Kaastra16}, i.e. \texttt{grouptype=optmin} with \texttt{groupscale=10} in the \texttt{ftgrouppha} command, and performed all spectral fitting with the Cash statistic \citep{Cash1979}.

\subsection{XMM-Newton} \label{subsec:xmm}
\textit{XMM-Newton} \citep{Jansen01} observed Ansky across three Director's Discretionary Time observations (ObsIDs 0935191401-0935191601) on June 27/June 29/July 1, 2024. As seen in Fig.~\ref{fig:lc_spec}, ObsID 0935191401 was taken well into the QPE decline phase; 0935191501 occurred entirely during quiescence; and 0935191601 was taken just after peak, catching the beginning of the QPE decline.

The data were retrieved from the \textit{XMM-Newton} Science Archive reduced using XMM SASv21.0.0 and HEASoft v6.33. The EPIC-pn instrument was operated in small window mode to mitigate against pile-up from the high count rates of the source. Source products were extracted from a circular region of 40'' radius, while the background was extracted from source-free circular region of 60'' radius. We retained events with PATTERN$\leq$4 (single and double events only) and discarded time intervals with a 10--12 keV count rate $\geq 0.4$ counts s$^{-1}$. This resulted in a net background-filtered exposure time of 87.4~ks: 28.2~ks in ObsID 0935191401, 25.1~ks in ObsID 0935191501, and 34.1~ks in ObsID 0935191601. We extracted spectra with the \texttt{evselect} command, without grouping the data (i.e. keeping 1 count per bin) and fit with the Cash statistic. For ObsIDs 0935191401 and 0935191601, we also extracted phase-resolved spectra by dividing the data into three equal-duration segments with \texttt{evselect}.

Despite the operation of the EPIC-pn camera in small window mode, pile-up remains a potential concern for the especially soft spectra like that of Ansky. We thus checked whether extracting source products in an annular region, with varying inner radii excised between 5''/10''/15'', significantly affected the spectral products. We found no significant changes among the different source regions. Moreover, we see similar spectral residuals in both \textit{NICER} and RGS, neither of which are affected by pile-up.

RGS data were reduced with the \texttt{RGSPROC} command, following the standard procedures and settings. We centered the extraction region on the position of Ansky and filtered background flares above 0.25 counts/sec, resulting in 34.0~ks net exposure for the RGS 1 spectrometer and 33.9~ks for RGS 2. Data from the two RGS instruments were binned by a factor of $3\times$ the native instrumental resolution and fit separately with a free normalization factor, and were then stacked to produce the overall spectrum shown in Fig.~\ref{fig:rgs_simult}. In ObsID 0935191501 the source flux is too low to produce a useful RGS spectrum, as the instrument is background-dominated at all energies. ObsIDs 0935191401/0935191601 are significantly brighter, containing 4497/10441 total RGS counts respectively. We carry out spectral analysis only on ObsID 0935191601 due to the superior SNR of that spectrum and the difficulty of constraining models with $\lesssim$10,000 counts. We chose not to stack with data from ObsID 0935191401, because the feature evolves rapidly. All RGS fitting was performed in the range 0.3--1.15 keV, as the spectrum is background-dominated at higher energies.

\section{Supplementary figures and tables} \label{subsec:suppl}
\begin{figure*}
    \centering
    \includegraphics[width=\textwidth]{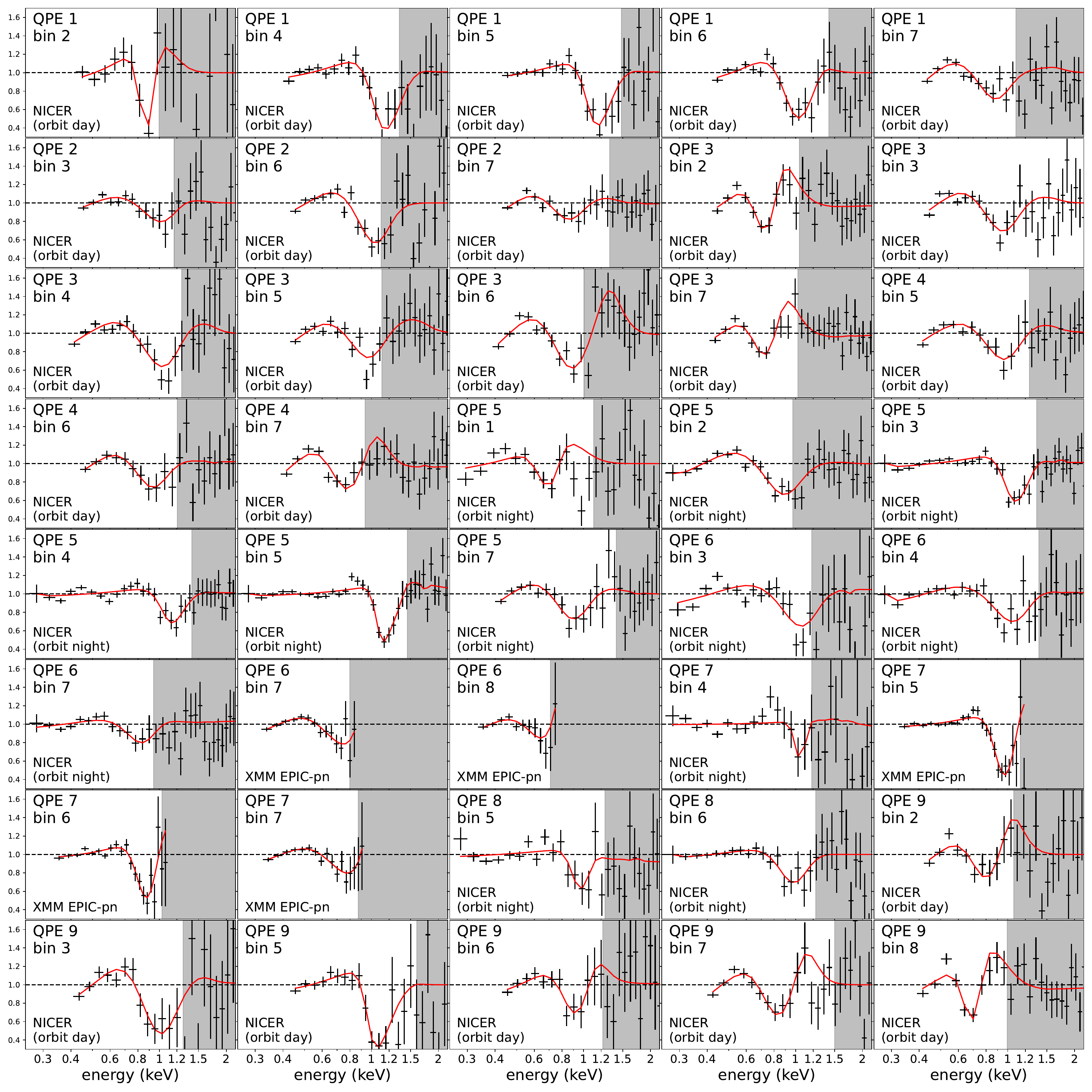}
    \caption{Data/model ratio for a blackbody-only model. The red line denotes the model with a gaussian absorption profile added, which significantly improves the fit in all cases. The gray shaded areas denote background-dominated energies, i.e. points where the \textit{NICER} model background or \textit{XMM} observational background greater than or equal to the blackbody source model.}
    \label{fig:ratio_plots}
\end{figure*}

In Fig.~\ref{fig:ratio_plots}, we show the data-to-model ratios for the blackbody-only models in all 35 \textit{NICER} spectra and all 5 \textit{XMM} EPIC-pn spectra used in Fig.~\ref{fig:evol}. We also show the effectiveness of a gaussian absorption profile in removing the residuals.

Most of the \textit{XMM} data taken during QPE 7 occured during a period of higher-than-average \textit{NICER} background contamination, which unfortunately prevented us from comparing simultaneous data from both instruments. The only \textit{XMM} bin with sufficiently high QPE flux and sufficiently low background for the Wien-tail residuals to be accessible to \textit{NICER} is QPE 6 bin 7, during which the two instruments appear to agree on the absorption feature.

\bibliography{refs}{}
\bibliographystyle{aasjournal}

\end{document}